\documentclass[letterpaper]{article} 
\usepackage{aaai24}  
\usepackage{times}  
\usepackage{helvet}  
\usepackage{courier}  
\usepackage[hyphens]{url}  
\usepackage{graphicx} 
\usepackage{natbib}  
\usepackage{caption} 
\usepackage{algorithm}
\usepackage{algorithmic}

\usepackage{newfloat}
\usepackage{listings}
\DeclareCaptionStyle{ruled}{labelfont=normalfont,labelsep=colon,strut=off} 
\floatstyle{ruled}
\newfloat{listing}{tb}{lst}{}
\floatname{listing}{Listing}
\nocopyright 

\title{ExploreGen: Large Language Models for \\Envisioning the Uses and Risks of AI Technologies}
\author{
    Viviane Herdel\textsuperscript{\rm 1}, Sanja \v{S}\'{c}epanovi\'{c}\textsuperscript{\rm 2}, Edyta Bogucka\textsuperscript{\rm 2}, 
    Daniele Quercia\textsuperscript{\rm 2,3}\\
}
\affiliations{
    \textsuperscript{\rm 1}Ben-Gurion University, Negev, Israel\\
    \textsuperscript{\rm 2}Nokia Bell Labs, Cambridge, UK\\
    \textsuperscript{\rm 3}Kings College London, Londond UK\\

}

\usepackage{bibentry}

\usepackage{booktabs}
\usepackage{array}
\usepackage{longtable}
\usepackage{arydshln}

\usepackage{etoolbox}
\usepackage{color}
\usepackage{soul}

\newbool{showComments}
\boolfalse{showComments}
\ifbool{showComments}{

}

\begin{document}

\maketitle

\begin{abstract}
Responsible AI design is increasingly seen as an imperative by both AI developers and AI compliance experts. One of the key tasks is envisioning AI technology uses and risks. Recent studies on the model and data cards reveal that AI practitioners struggle with this task due to its inherently challenging nature. Here, we demonstrate that leveraging a Large Language Model (LLM) can support AI practitioners in this task by enabling reflexivity, brainstorming, and deliberation, especially in the early design stages of the AI development process.
We developed an LLM framework, \emph{ExploreGen}, which generates realistic and varied uses of AI technology, including those overlooked by research, and classifies their risk level based on the EU AI Act regulation. We evaluated our framework using the case of Facial Recognition and Analysis technology in nine user studies with 25 AI practitioners. Our findings show that \emph{ExploreGen} is helpful to both developers and compliance experts. They rated the uses as realistic and their risk classification as accurate (94.5\%). Moreover, while unfamiliar with many of the uses, they rated them as having high adoption potential and transformational impact. 
\end{abstract}

\section{Introduction}\label{sec_intro}
In today's fast-paced tech world, balancing innovation with responsibility is essential \cite{sraml2022innovation,owen2019responsible}. As Artificial Intelligence (AI) spreads across areas like healthcare and finance, it is crucial to understand its uses and potential risks relating, e.g., to data privacy, security, and fairness \cite{davenport2019potential,goodell2021artificial,dignum2019responsible,tahaei2023human}. 
Business developers and engineers seek opportunities to employ the latest AI trends ahead of their competitors \cite{phaal2004technology}, while researchers take part in a similarly fast-paced environment to publish their latest AI discoveries. In both roles, these AI practitioners are faced with increased need to envision potential uses, as well as risks and benefits of the technologies they are developing, and to produce AI impact assessment  reports \cite{stahl2023systematic}. 
Given the increasing number of AI regulations \cite{smuha2021race}, AI compliance experts also face the task of supporting their colleagues in assessing the regulatory risks and compliance of AI technologies. The process of cataloging AI uses and associated risks is both challenging and time-consuming \cite{moraes2021smile,liang2024s,hassel2023governing}.
Recent research shows that AI developers struggle with detailing uses and impacts for model cards \cite{liang2024s} and data cards \cite{yang2023navigating}, as well as for the broader societal impacts sections now mandated by some of the top AI conferences \cite{nanayakkara2021unpacking,prunkl2021institutionalizing,ashurst2022ai}.
Recommendations to support AI practitioners with envisioning the impacts of their technology include encouraging reflexivity, including constructive and data-driven deliberation \cite{ashurst2022ai,prunkl2021institutionalizing,yang2023navigating}.

Our research responds to this challenge by exploring the use of Large Language Models (LLMs) to generate AI technology uses and their risk assessments based on the EU AI Act \cite{EUACT2023}. This aims to support AI practitioners during the initial phases of the AI design process, including reflexivity, brainstorming, and deliberation.
While LLMs have demonstrated utility in diverse applications \cite{gilardi2023chatgpt,CHI_AI_CHAINS,dowling2023chatgpt,byun2023dispensing}, their suitability for two specific tasks—identifying potential uses of a given AI technology and conducting legal risk assessments of its uses—remains an open question. Our aim is not to produce an exhaustive list of uses for a given AI technology, nor to provide a definitive risk classification. Instead, we aim to investigate whether LLMs can generate outputs of sufficient quality to support AI practitioners in envisioning the impacts of their technology, particularly focusing on \emph{less well-researched uses}.
On one hand, LLMs might generate unrealistic use cases or ones that practitioners are already familiar with. On the other hand, the extent to which LLMs can accurately map legal regulations to specific AI uses, if at all, is yet to be substantiated. 

\begin{figure*}[!htb]
\centering
\includegraphics[width=\linewidth]{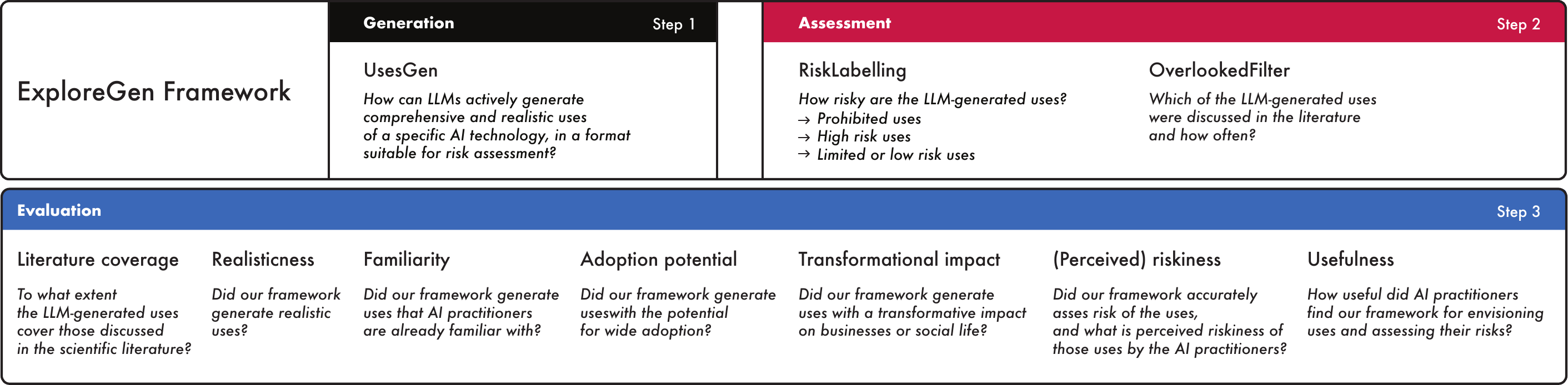}
\caption{Our methodology consists of three steps. In the first two steps, \emph{ExploreGen} performs \emph{(i)} \textit{generation (UsesGen)} of various uses for a given AI technology, and their \emph{(ii)} \textit{assessment (RiskLabeling, OverlookedFilter)} in terms of the risks based on the EU AI Act, and determining whether they are discussed or overlooked in previous literature. In the last step \emph{(iii)}, we did the \textit{evaluation} of the generated uses and their risk classification, including the realisticness of the uses, risk assessment accuracy, and usefulness for AI practitioners in envisioning the impacts of AI technology.}
  \label{fig:teaser}
\end{figure*}

This paper aims to evaluate LLMs for these specific goals. We explored them using OpenAI's GPT-4 \cite{openai2023gpt4}, making two main contributions (Figure~\ref{fig:teaser}):

\begin{enumerate}
\item We designed an LLM framework (\emph{ExploreGen}) incorporating novel prompt elements—a set of curated \emph{domains} to generate a variety of uses, and \emph{risk concepts} proposed by \citet{Golpayegani2023Risk}, framing each use along these concepts for risk assessment (\emph{UsesGen}). \emph{UsesGen} classifies generated uses into realistic (existing and upcoming) and unlikely (hallucinations) with Chain-of-Thought (CoT) reasoning \cite{NEURIPS2022_9d560961}, retaining only \emph{realistic} ones. These uses are then classified into prohibited, high-risk, and limited or low-risk categories according to the EU AI Act (\emph{RiskLabelling}). Additionally, we processed 3M Semantic Scholar papers, to uncover $\sim$12\% among the identified uses, which were overlooked by the scientific literature (\emph{OverlookedFilter}).

\item  
Using Facial Recognition and Analysis (FRA) technology as a use case, we \emph{evaluated} our framework by assessing six aspects: \emph{(I)} whether it generates realistic uses, \emph{(II)} literature coverage of the generated uses, \emph{(III)} familiarity of AI practitioners with these uses, \emph{(IV)} adoption potential, \emph{(V)} transformational impact, and \emph{(VI)} accuracy of risk classification and perceived riskiness by AI practitioners.

To perform the evaluation, we conducted a scoping literature review, and nine user studies with 25 AI practitioners (12 AI developers and 13 AI compliance experts). We found that \emph{UsesGen} generated realistic uses, covering 96\% of the literature uses identified through the scoping review \emph{(I-II)}. AI practitioners reported low familiarity with the uses, especially the overlooked ones \emph{(III)}. They considered the uses somewhat to very likely to be adopted \emph{(IV)} and to have a high transformational impact on business operations or people's lives \emph{(V)}. Compliance experts found that \emph{RiskLabelling} correctly classified the risk of uses based on the EU AI Act with a 94.5\% accuracy. Although over 50\% of the FRA uses were classified as high risk or prohibited, AI developers, who were not presented with the classification, perceived most uses as only slightly risky for society and not at all for the environment. Lastly, thematic analysis of open-ended responses during in-person interviews revealed that both AI developers and compliance experts found \emph{ExploreGen} helpful for ideation, brainstorming, and deliberation of AI uses and their risks and benefits. Compliance experts found it directly useful, while developers recommended adjustments to better suit their needs.

\end{enumerate}

\section{Background \& Related Work}\label{sec_background_rw}
First, we present background on assessing impacts of AI technology, followed by a glimpse on the emergent AI regulations, and we finish with prior work leveraging LLMs for various tasks.

\subsection{Assessing Impacts of AI Technology}\label{sec_rw_impact_assessment}
AI impact assessments (AIIAs) are recommended as a tool to recognize both the beneficial and adverse effects early in the AI technology development process, aiming to predict and evaluate the impact that new digital technologies have on all stakeholders. \citet{stahl2023systematic} reviewed literature and identified 38 proposed AIIAs, including DataSheets for Datasets \cite{gebru2021datasheets} and methods inspired by environmental impact assessments \cite{calvo2020advancing}. However, despite the proliferation of proposed AIIAs, developer teams often encounter difficulties initiating AI impact assessments \cite{buccinca2023aha} and require additional guidance throughout this process \cite{wang2023designing}. 

An important challenge faced by AI practitioners when performing AI impact assessments is mapping the intended and unintended AI uses \cite{liang2024s,yang2023navigating,prunkl2021institutionalizing}. For example, recent research on 32K model cards posted on the HuggingFace platform \cite{liang2024s} shows that while most cards detail \emph{Training Information}, sections on \emph{Intended Uses} and \emph{Bias, Risks, and Limitations} have lower completion rates (17-23\%). Similarly, \citet{yang2023navigating} found that in Data Cards also hosted on HuggingFace, the section on \emph{Considerations for Using the Data} receives the lowest proportion of content (only 2.1\% of the card's text length). 

As another means of reflecting on potential positive and negative consequences of AI models, broader societal impacts are introduced as a requirement by leading AI conferences (e.g., the Conference on Neural Information Processing Systems (NeurIPS)) \cite{nanayakkara2021unpacking}. However, researchers also struggle with filling in such sections due to the inherently difficult nature of the task and high opportunity costs \cite{prunkl2021institutionalizing}. 

Conventional methods to understand the uses and scope of AI technology include systematic and scoping reviews, which are useful for mapping fields of study \cite{Peters2015Scoping,loncar2019literature}. For instance, \citet{moraes2021smile} combined literature review with news media research to unveil FRA applications in (semi-)public spaces in Brazil and the associated risks. Similarly, \citet{hupont2022landscape} reviewed scientific papers and company portfolios to identify 60 facial processing applications, which were then assessed for risk level according to the EU AI Act. However, these methods, while insightful, are resource-intensive, demanding both time and expertise \cite{Arksey2005}.

Moreover, even when the uses of AI are known, they can bring unanticipated challenges, from privacy and security issues \cite{li2023trustworthy,chi_air_risk_children_apps} to distorting human beliefs \cite{kidd2023ai}, excessive dependence that could diminish crucial human skills \cite{byun2023dispensing,chi_reliance_ai}, and negative environmental impacts \cite{rillig2023risks}, as well as impacts on human rights and society \cite{mantelero2022}. Anticipating such challenges and broader, systemic impacts of technology remains a significant challenge for AI practitioners \cite{prunkl2021institutionalizing,yang2023navigating,weidinger2023sociotechnical}.

\subsection{Regulating AI}\label{sec_rw_regulation}
The pervasiveness of AI, along with the potential risks discussed above, has intensified calls for regulatory oversight \cite{tahaei2023human, borenstein2021emerging}. The first binding regulatory response is the European Commission's AI Act \cite{EUACT2023}, which aims to balance fostering innovation with protecting rights and societal values. The Act covers a spectrum from low-risk to prohibited AI applications, prohibiting those that can harm individuals or manipulate behaviors, such as social scoring by public authorities. It also allows for extending the scope of prohibited and high-risk uses, recognizing that AI regulations must evolve alongside technology \cite{nature2023aireg}. 

Other regulatory frameworks include the US Office of Science and Technology Policy (OSTP) Blueprint for an AI Bill of Rights, China's Interim Measures for the Management of Generative AI Services, and the UK's pro-innovation approach to AI regulation.


To sum up, the dynamic nature of AI poses a challenge in its impact assessment, particularly in identifying its myriad uses and ensuring thorough risk assessments. We propose to leverage LLMs to partly tackle these challenges. 

\subsection{Large Language Model Applications}\label{sec:llm}
LLMs have already demonstrated their usefulness in a variety of tasks. These range from text annotation \cite{gilardi2023chatgpt} to assisting with creative and argumentative writing \cite{chi_llm_coauthor} and potential for providing help for mental health issues \cite{sharma2023cognitive}. LLMs offer insights that surpass general public knowledge \cite{gilardi2023chatgpt}, show promise in human-AI co-creation processes \cite{CHI_AI_CHAINS,chi_llm_coauthor}, {brainstorming assistants \cite{lukowicz2023interacting,bouschery2024artificial}}, and have the potential to support \emph{interpreting regulatory texts} \cite{zheng2023llm,cui2023chatlaw}. 

To achieve the desired output from LLMs, it is important to employ best practices in prompt engineering, such as Chain-of-Thought reasoning, using appropriate roles, and providing cues and examples to guide the model's output \cite{CHI_AI_CHAINS,OpenAIGuide}. However, LLMs also introduce their own AI risks, including biases associated with the training data \cite{luccioni2024stable} and hallucinations \cite{mittelstadt2023protect}, which need to be carefully considered in each application.

\section{Methodology}\label{sec_methods}
For our framework's development and assessment, we focused on Facial Recognition and Analysis (FRA), a well-established yet controversial technology due to its known risks \cite{zhang2021facial,mcclurg2007face}, and a contentious topic during the development of the EU AI Act \cite{hupont2022landscape}.

\subsection{Designing ExploreGen}
\label{sec_ExploreGen}
We selected GPT-4 due to its top-ranking performance, as shown in leaderboards~\cite{leaderboard2}.

\mbox{ } \\
\noindent
\textbf{Generating Uses (UsesGen).} To generate a list of various uses  (Figure \ref{fig:teaser}, Framework, Step 1), we specified five elements in UsesGen (Appendix, Figure \ref{Fig:Generation}): system role, instructions, risk concepts, definitions of being realistic, domains, and examples. 

The \emph{system role} has been shown to improve the quality of the output, as it allows to generate content from specific perspectives \cite{giray2023prompt}. We assigned the role of a ``\emph{Senior [Technology X] Specialist and Evaluator}'' and described its main tasks as \textit{``reviewing, and cataloguing the diverse applications and use cases of [Technology X] across multiple domains, and conducting exhaustive research and analysis''}.

We then followed with the three-part \emph{instruction}: (i) to create a comprehensive and self-explanatory JSON (JavaScript Object Notation) list detailing particular use cases or applications of [Technology X], (ii) to provide precise descriptions for each concept, and (iii)
to categorise the LLM-generated uses into 1) \emph{already existent}, 2) \emph{upcoming}, and 3) \emph{unlikely}, along with a one-sentence justification for each use categorization (enacting the \emph{CoT reasoning}).

We asked for each use to be generated along the five concepts proposed by \citet{Golpayegani2023Risk}:
\begin{enumerate}
    \item \textit{Domain}: ``The area or sector the AI system is intended to be used in'' (e.g., education). 
    \item \textit{Purpose}: ``The objective that is intended to be accomplished by using an AI system'' (e.g., attendance tracking).
    \item \textit{Capability}: ``The capability of the AI system that enables the realisation of its purpose and reflects the technological capability'' (e.g., identify students' faces and match them with database).
    \item \textit{AI user}: ``The entity or individual in charge of deploying and managing the AI system, including individuals, organisations, corporations, public authorities, and agencies responsible for its operation and management'' (e.g., schools).
    \item \textit{AI subject}: ``The individual directly affected by the use of the AI system, experiencing its effects and consequences. They interact with or are impacted by the AI system's processes, decisions, or outcomes'' (e.g., students).
\end{enumerate}

To aid the realisticness categorisation, we also provided the \emph{definitions} of the three categories of being realistic. Already existent uses were defined as currently implemented and well-established uses. Upcoming uses were defined as being under current development, being researched, or subject to discussions without being implemented or being severely limited in practice due to various reasons. Lastly, unlikely uses, introduced to capture hallucinations, lack value, usability, applicability, or practicality, or are deemed unnecessary, impossible, incoherent, or unrealistic. 

To further guide UsesGen we requested the AI technology uses across a broad set of \emph{domains}. Without such a request, the uses generated by the LLM would encompass the most common and well-known FRA uses, since LLMs suffer from exposure bias \cite{CHI_AI_CHAINS}. The domains served as a \emph{cue} in our prompt. Our procedure for listing a broad set of domains was as follows. First, domains were derived from the EU AI Act's Annex III (e.g., ``Education and vocational training''), along with 32 domains that were not explicitly listed but were mentioned in the EU AI Act text or its Amendments (e.g., ``Social Media'' from Amendment 51 stating: \textit{``The indiscriminate and untargeted scraping of biometric data from social media [...] add to the feeling of mass surveillance [...]''}).
Moreover, we derived additional domains from a focus group using a think aloud protocol ($N$=8) to ensure capturing all significant domains beyond the EU AI Act. The session was with our research group (3F, 5M, mean age: 31.8, $SD$: 6.74, range: 22-45). We used a Miro board and asked the participants to think of domains that affect their lives along the five levels of the Social-Ecological Model \cite{Shelley2015Upending}: individual, interpersonal, institutional, community, and public policy. This resulted in an additional 6 domains that were not yet covered by the previous list of 40 domains, resulting in the final 46 diverse domains (Appendix (B)). 

To complete the prompt, we carefully crafted five \emph{examples} (employing \emph{few-shot} learning \cite{brown2020language}) striking a balance between providing a diverse range of examples and keeping the prompt at a manageable context length \cite{liu2024lost}. 
The output also requires the label for the realisticness of the use.
For example, ``FRA for medical diagnosis'' was categorised as an \emph{upcoming} use, along with the justification saying that it \emph{has the potential to revolutionise healthcare, yet successful integration depends on resolving privacy, regulatory, and trust-related issues.} 
We placed the examples section at the end of the prompt, as examples can not only illustrate the desired input-output relationships, but also aid the model's context comprehension and response expectations \cite{brown2020language}.

\mbox{ } \\
\noindent
\textbf{Assessing the Risk of Generated Uses (RiskLabelling).} 

To enable risk assessment as per the EU AI Act (Figure \ref{fig:teaser}, Framework, Step 2), we specified five elements (Appendix, Figure \ref{Fig:Risk}): system role, instructions, legal documents, placeholder for a list of uses, and output structure. 

We started the prompt by selecting the \emph{system role} of an \textit{``Experienced Judge who works in the field of AI technology regulation''}, and described the role further: \textit{``You are thoughtful, decisive, experienced and conscientious. You have access to the entirety of the EU AI Act''}.

We then provided the \emph{instructions} to classify the uses by utilising the \emph{CoT reasoning} by requesting to first expand the concise FRA use into a description of a hypothetical AI system that employs it. 
We then followed with the requests to consider the EU AI Act and its  amendments \cite{EUACT2023} provided in \emph{input}, and to classify the system as ``prohibited'', or ``high risk'', or, otherwise, as ``limited or low risk''.

The prompt was then provided with the \emph{placeholder} for AI technology uses for which the risk assessment should be performed. 

Finally, we requested the \emph{output structure} of the risk classification to encompass:
\begin{enumerate}
    \item \textit{Description}: Provides a clear understanding of the intended use of the AI system.
    \item \textit{Classification}: Outcome of the classification which can be either prohibited, high risk, or limited or low risk.
    \item \textit{Relevant Text from the Act}: If applicable, a quote from the EU AI Act is included, along with a relevant amendment or section to provide legal context.
    \item \emph{Reasoning}: Explanation that rationalises the specific risk classification of the inputted AI use.
\end{enumerate}

\begin{table}[htb!]
\centering
\caption{LLM-generated uses overlooked by the research literature. For full details of these uses, see Appendix, Table \ref{tab:llmuses}.}
\label{tab:overlooked_uses}
\footnotesize
\begin{tabular}{l}
\hline
\textbf{Use ID. Use Description} \\ \hline
27. Validate remote worker identity online. \\ 
52. Recognize customers, tailor services. \\ 
68. Identify watchlisted individuals at borders. \\ 
69. Verify asylum seeker identities. \\ 
70. Prevent voter fraud via identity verification. \\ 
80. Authenticate energy facility personnel access. \\ 
83. Verify military personnel identities. \\ 
84. Identify threats in crowds by military. \\ 
88. Identify citizens for personalized services. \\ 
91. Secure embassies by identifying visitors. \\ 
98. Authenticate emergency responders’ identities. \\ 
104. Verify cargo access by authorized personnel. \\ 
108. Control access to restricted urban areas. \\ 
114. Verify access to protected environmental areas. \\ 
118. Identify illegal loggers. \\ 
120. Verify access to climate-sensitive areas. \\ \hline
\end{tabular}
\end{table}

\mbox{ } \\
\noindent
\textbf{Assessing the Literature Coverage of Generated Uses (OverlookedFilter).} To assess which of the LLM-generated uses were discussed in the literature (Figure \ref{fig:teaser}, Framework, Step 2), and possibly uncover overlooked ones by the literature, we collated all the 200M papers from Semantic Scholar's May 2023 dump.\footnote{api.semanticscholar.org/api-docs/datasets} We then filtered the papers to those being written in English, and having both the title and abstract fields available, resulting in 3M papers.

Next, we embedded the \texttt{title + abstract} field for each of the articles, as well as the description of each of the LLM-generated use using \emph{all-mpnet-base-v2} sentence-transformers \cite{reimers2019sentence} model.\footnote{huggingface.co/sentence-transformers/all-mpnet-base-v2} This model is trained using a self-supervised contrastive learning, by fine tuning the pretrained \emph{microsoft/mpnet-base} model on above 1 billion sentences. Upon pairing each use with the paper with the maximum similarity of their embeddings, we then manually explored which similarity threshold will yield use-paper pairs such that the paper's abstract indeed discusses the use. We explored $ \{95^{th}, 99^{th}, 995^{th}, 999^{th}\}$  percentile thresholds, until we concluded that  the $999^{th}$ percentile one yielded $3,295$ papers, which indeed discussed paired FRA uses.   

The top frequent venues in which these papers are published include: arXiv.org, International Journal for Research in Applied Science and Engineering Technology, IEEE International Conference on Systems, Man and Cybernetics, ACM Multimedia, Interspeech, PLoS ONE, IEEE/ACM International Conference on Human-Robot Interaction, and Computer.
The most commonly discussed uses are: \emph{Secure access control, use \#1} discussed by 291 articles, \emph{Detecting driver fatigue through facial analysis, use \#134} discussed by 251, and \emph{use \#60, Using diverse facial data to refine algorithms}, discussed by 189 articles. 

\subsection{Evaluating ExploreGen}
\label{sec_evaluation_methods}
This section outlines the process of evaluating our ExploreGen framework (Figure \ref{fig:teaser}, Framework, Step 3).
The goal of our framework was to generate realistic uses of a given AI technology, such that AI practitioners are not familiar with all of them, and to accurately classify their risks based on the regulation. Moreover, the generated uses should exhibit potential for adoption and transformational impact.

To ascertain the effectiveness of the framework at meeting this goal, our evaluation ought to answer seven questions:
\begin{enumerate}
    \renewcommand{\labelenumi}{\Roman{enumi}.}
    \item \emph{Literature coverage.} To what extent the generated uses cover those discussed in the scientific literature?
    \item \emph{Realisticness.} Did our framework generate realistic uses?
    \item \emph{Familiarity.} Did our framework generate uses AI practitioners are familiar with?
    \item \emph{Adoption potential.} Did our framework generate uses that have a potential for adoption?
    \item \emph{Transformational impact.} Did our framework generate uses that have a transformation impact?
    \item \emph{(Perceived) riskiness.} Did our framework accurately asses risk of the uses, and what is perceived riskiness of those uses by the AI practitioners?
    \item \emph{Usefulness.} How useful did the AI practitioners find our framework in assisting with their tasks of envisioning AI uses and assessing associated risks?
\end{enumerate}

\subsubsection{Metrics.}
We then defined six quantitative and one qualitative metric to answer these questions. 

The first metric assessed the \emph{coverage} of the generated use cases in relation to those discussed in the literature. It was measured as the percentage of matches with the ground truth (GT), which we derived from a scoping review of FRA use cases (Appendix C). Two authors independently conducted a manual assessment, categorizing each generated use case as either matching or not matching the ground truth list. 

The second metric assessed the \emph{realisticness} of the generated uses. We measured it by calculating the agreement between the realism labels assigned by the LLM and those given by the participants in the user study. 

The third metric assessed participants' \emph{familiarity} with the generated uses. It was measured through a question: \emph{``How frequently do you encounter references to this use in your professional life?''} evaluated on a 7-point Likert scale from `rarely' to `always'. 

The fourth metric assessed practitioners' perceptions about the real-life \emph{adoption potential} of the LLM-generated uses. It was measured through a question: \emph{``How likely it is that this use will be widely adopted in the near future?''} evaluated on a 7-point Likert scale from `very unlikely' to `very likely'. 

The fifth metric assessed AI practitioners' perceptions of the potential \emph{transformational impact} of the LLM-generated use cases. It was measured by asking, \emph{``How likely is it that this use will fundamentally change the way businesses operate or people live?''}. Participants rated this on a 7-point Likert scale from `very unlikely' to `very likely'. 

The sixth metric assessed AI practitioners' perceptions of the \emph{riskiness} of the use cases in terms of their potential societal and environmental adverse impacts. It was measured by asking both AI developers and compliance experts to answer how risky do they consider the use \emph{``for society as a whole''} as well as \emph{``for the environment''}. These two questions were rated on a 7-point Likert scale from `not risky at all' to `unacceptably risky'. Additionally, to validate \emph{RiskLabelling}'s classification outputs, we provided the compliance experts with both the classification and the LLM's justification and measured their agreement. If they disagreed with the classification, they could select the correct classification (including the option of `insufficient information to assess the use'). If they disagreed with the justification, they could provide their own reasoning.

The last, seventh metric was about the \emph{usefulness} of our framework, captured through three open-ended questions: \emph{``How useful is this framework for envisioning uses of technology?''}, \emph{``How useful is this framework for understanding the risks and benefits of each use?''}, and \emph{``At what stage in your assessment process would you use this framework?''} .

\subsubsection{Setup.}
To derive the first metric \emph{(literature coverage)}, we performed a scoping review. To derive the remaining six metrics \emph{(realisticness, familiarity, adoption potential, transformational impact, perceived riskiness, usefulness)}, we conducted nine user studies with 25 AI practitioners in total (12 AI developers, and 13 AI compliance experts).

\mbox{ } \\
\noindent
\textbf{Scoping Review.}
To obtain a list of a FRA uses discussed in the literature, we performed the scoping review in accordance with the 5-stage guidelines \cite{Arksey2005}:  
\begin{figure}[t]
\centering
\includegraphics[width=0.3\textwidth]{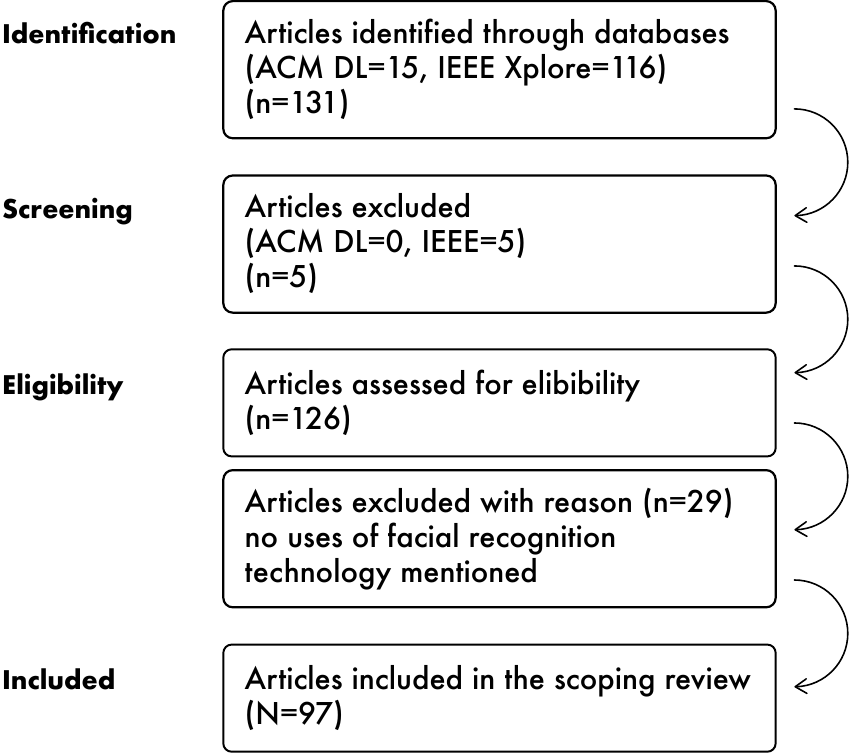}
\caption{\emph{The scoping review:} identification, screening, and assessment for eligibility of articles. Starting with 131 initial papers identified, a total of 97 were included. From these papers, 75 unique FRA uses were identified (Appendix C).} 
\label{Fig:Prisma}
\end{figure}

\begin{enumerate}
\item \emph{Identifying research questions.} RQ: ``What are the documented, researched, or proposed uses of FRA as found in the literature?''

\item  \emph{Identifying relevant articles.}
In consultancy with the research team, we selected the ACM Digital Library (https://dlnext.acm.org) and IEEE Xplore (http://ieeexplore.ieee.org) as our databases, which correspond to the main Computer Science and Engineering digital libraries, likely to cover a broad spectrum of research on FR technology. We used the following Query (Q) = [[Title: ``face recognition''] OR [Title: ``facial recognition'']] AND [Full Text: ``use case*''] (Figure \ref{Fig:Prisma}, \textit{Identification}). 

\item  \emph{Selecting articles.}
We included peer-reviewed articles as well as larger scholarly works, such as extended abstracts (e.g., posters and demos) and workshop papers. All selected works are referred to as \textit{articles}. For all identified articles, we applied the following inclusion criteria: (1) written in English and (2) describing, studying, or envisioning at least one use of facial recognition technology.
In the initial search, 131 articles were identified from the ACM and IEEE databases. As no duplicates were found, all 131 articles were screened based on titles and abstracts. Five articles were removed as they did not discuss an FRA use (Figure \ref{Fig:Prisma}, \textit{Screening}). Subsequently, 126 articles were assessed for eligibility based on their full text (Figure \ref{Fig:Prisma}, \textit{Eligibility}), resulting in a final selection of 97 relevant articles (Figure \ref{Fig:Prisma}, \textit{Included}). The lead author performed the article selection process.

\item  \emph{Charting the data.} 
The lead author began reading the articles and simultaneously developed a coding system for the FRA uses described, studied, and envisioned in the articles. As the lead author read the articles, they color-coded the FRA uses and extracted them. Each time a new FRA use was identified, it was added to the coding system. Any ambiguities—though rare due to the straightforward nature of the FRA uses mentioned—were discussed and resolved between the lead and second author.

\item \emph{Collating, summarising, and reporting results.}
The scoping review resulted in 97 articles from which we identified 75 unique uses of FRA, listed in Appendix C.
\end{enumerate}

\mbox{ } \\
\noindent
\textbf{User Studies with AI Practitioners.}
We conducted seven in-person studies involving 3 AI developers (30 minutes each) and 4 AI compliance experts (45 minutes each), complemented by two online studies on Prolific: one with 9 AI developers, and another one with 9 AI compliance experts. 

The in-person studies consisted of four steps. First, we asked participants about their current practices and challenges in envisioning AI technology uses and their associated risks. Second, we presented an interactive list of 138 uses and tasked them with selecting one project that balances being interesting to develop and adhering to the company's code of conduct (Figure \ref{Fig:studies-materials}A), followed by questions on the usefulness of this list for envisioning technology uses and understanding the risks and benefits. Third, we presented 16 interactive assessment cards for overlooked uses and tasked them with annotating the uses for realisticness, familiarity, adoption potential, transformational impact, and perceived riskiness (Figure \ref{Fig:studies-materials}B). AI compliance experts also evaluated the \emph{RiskLabelling} classification and justification, making corrections if necessary (Figure \ref{Fig:studies-materials}C). This allowed us to compare perceived use riskiness between developers and compliance experts. Finally, we asked participants about the framework's usefulness for envisioning technology uses, understanding risks and benefits, and identifying the stage in their assessment process where they would use this framework. Each of the 16 uses was annotated by 7 different AI practitioners: 3 AI developers and 4 AI compliance experts. 

The online studies used a custom web-based survey consisting of five pages. The first page outlined the study's description and tasks for crowdworkers: read the definitions of `risky' uses and annotate each use for realism, familiarity, adoption potential, transformational impact, and perceived riskiness. AI compliance experts were also asked to agree or disagree with the \emph{RiskLabelling} classification and justification, and make corrections if necessary. The second page provided definitions of risky uses according to the EU AI Act. The third and fourth pages presented assessment cards for 46 uses (23 per page) with input boxes for annotations (Figure \ref{Fig:studies-materials}A,B). The final page included a confirmation note and redirected participants to Prolific. Each of the 138 uses was annotated by 6 different AI practitioners: 3 AI developers and 3 AI compliance experts. 

To ensure response quality, we conducted two attention checks during the studies and implemented two deliberate survey design features. First, after reading task instructions, participants encountered one of the two attention-check sentences: \emph{``When asked for your favorite color/city, you must select ``Blue/Rome''}. We also included one prohibited use labelled as ``low risk'' with a false justification mimicking text from the EU AI Act. Participants had to correctly respond to two out of these three checks. Second, we disabled pasting from external sources and editing previous responses to ensure original and thoughtful answers.

\subsubsection{Participants.}
For our studies, we recruited participants and surveyed them across two cohorts: \emph{a)} AI developers and \emph{b)} compliance experts.

For the in-person studies, we recruited participants through an internal mailing list at a large tech company, and our professional networks. We asked for individuals currently developing AI systems using machine learning, computer vision, and image recognition. To recruit AI compliance experts, we sought individuals familiar with the EU AI Act, experienced in reviewing AI use cases, and involved in at least one ongoing AI impact assessment project.

For the online studies, we recruited participants from Prolific, controlling for their roles in the organization, the frequency of AI use in their jobs, fluency in English, and geographic location. To recruit AI developers, we selected participants who likely contribute to developing AI systems as part of their software engineering roles, using AI daily. To recruit compliance experts, we looked for participants likely involved in revising AI systems as part of their legal roles, using AI at least 2-6 times a week. We limited our participant pool to individuals residing in the European Union. All Prolific participants were paid an average of \$12 USD/hour. 

\noindent\textbf{Analysis.} We performed both quantitative and qualitative analyses. 
For the quantitative analysis, we measured the frequencies across six metrics: coverage, realisticness, familiarity, adoption potential, transformational impact, and perceived riskiness. For the qualitative analysis, we thematically analyzed responses to open-ended questions \cite{saldana2015coding, miles1994qualitative, mcdonald2019reliability, braun2006thematic} to understand factors influencing the framework's usefulness for envisioning technology uses, assessing risks and benefits, and determining the appropriate  assessment stage for its application.

\section{Evaluation Results}
\label{sec_results}


\begin{figure*}[!htb]
\centering
\includegraphics[width=0.74\linewidth]{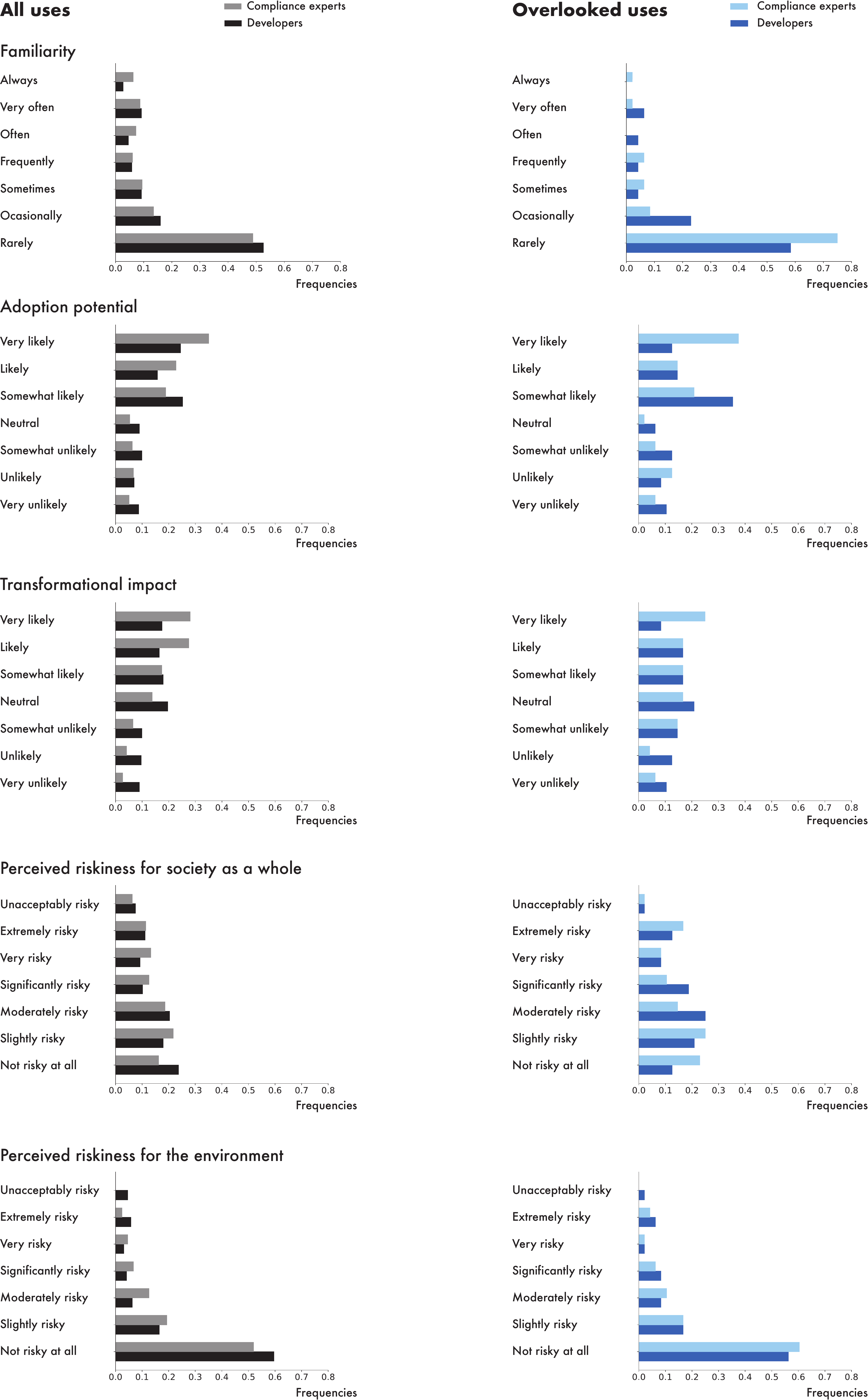}
\caption{Evaluation results for the five quantitative metrics: familiarity with the use, its adoption potential, transformational impact, and perceived riskiness for society as a whole and for the environment.}
\label{fig:eval}
\end{figure*}

\emph{UsesGen}, using FRA technology as input, generated 138 uses listed in Appendix D, Table \ref{tab:llmuses}. According to its own realisticness label, 8 (6\%) of the uses were deemed unlikely (e.g., \emph{FRA to track the carbon footprint of individuals, use \#119}, as it is unlikely to be adopted, and \emph{detecting plant diseases and pest infestations, use \#50}, as it does not employ the capabilities of FRA).

\emph{RiskLabelling} classified 10 (7\%) uses as prohibited, 66 (48\%) as high risk, and 62 (45\%) uses as limited or low risk. Example \emph{RiskLabelling} outputs for one use per each class are shown in Appendix, Table \ref{tab:riskex}.

\emph{OverlookedFilter} identified 16 out of the 138 LLM-generated uses that were not discussed in any of the 3 million Semantic Scholar papers we analyzed. These uses, which we term overlooked, are presented in Table \ref{tab:overlooked_uses}. This indicates that while these uses are likely mentioned in news, press, or social media (and thus included in the LLM training data), they have not yet been the focus of in-depth scientific research.




\vspace{0.3em}
\noindent
\textbf{I. Literature coverage.}
The uses were expressed differently between the GT list (Appendix C) and the LLM-generated list (Appendix D, Table 4). In the GT list, they are written as single sentences mainly describing the purpose, whereas in the LLM-generated list, they always follow a structured format based on the 5 risk concepts (e.g., AI domain, AI user). Therefore, we employed a relaxed matching approach, allowing us to count two uses with different levels of generality as a match (e.g., \emph{detect fatigue in individuals, GT-use \#69} was matched with \emph{improving driver safety by detecting driver fatigue through facial analysis, use \#134}).

The LLM-generated list covered 96\% of the literature-derived GT uses with the only 3 GT uses not found in the LLM-generated list being: \emph{Provide real-time information about visitors in high-profile buildings, GT-use \#5}, \emph{Help people recognise faces by using smart glasses to display names and social network activities of identified people, GT-use \#72}, and \emph{Facilitate tourists in meeting new people, GT-use \#74}.

Given the relaxed approach we applied, the high matching rate between the two lists reflects the LLM-generated list's scoping coverage of various uses discussed in the literature rather than comprehensively covering all possible uses. Given the many contexts for each use (e.g., various subjects, domains, or locations), comprehensive coverage is practically unattainable.


\vspace{0.3em}
\noindent
\textbf{II. Realisticness.}
After excluding the 8 uses labeled by the LLM itself as unrealistic, the majority agreement across the participants in different user studies was that the remaining 130 uses were all realistic. Of these, 91 uses (70\%) were labeled as already existing, and 39 (30\%) as upcoming (e.g., \emph{recognizing signs of distress or confusion for elderly care assistance, use \#6} and \emph{facilitating non-verbal communication by interpreting facial expressions and gestures for non-verbal individuals, use \#77}).

The analysis of unrealistic uses revealed that some domains were more prone to hallucination, such as "Agriculture and Farming" or "Environment and Sustainability." Given that FRA has fewer applications in these domains, asking the LLM to generate uses in these areas led to hallucinations. These domains were included because they are mentioned in the EU AI Act and hold potential significance for other AI technologies (e.g., Earth Observation), where they might not lead to hallucinated uses.

\vspace{0.3em}
\noindent
\textbf{III. Familiarity.}
As shown in Figure \ref{fig:eval}, both AI developers and compliance experts demonstrated low familiarity with the uses produced by \emph{UsesGen}. Over 50\% (48\%) of these uses were reported by developers (compliance experts) as rarely encountered in their professional lives. For the overlooked uses, developers reported rarely encountering 60\% of these, while compliance experts reported rarely encountering even 75\%. The chi-squared test results confirmed that the distributions of familiarity scores significantly differ between all uses and overlooked uses, validating the ability of our \emph{OverlookedFilter} to identify less well-known and understudied uses. The distribution of familiarity scores did not differ statistically significantly between the cohorts of AI developers and compliance experts.

\vspace{0.3em}
\noindent
\textbf{IV. Adoption potential.}
AI developers thought that most of the uses are `somewhat likely' ($\sim$27\% of the uses) or `very likely' ($\sim$25\% of the uses) to be adopted, though the ratio of the `very likely' ones was smaller for the overlooked uses ($<$15\% of the uses). Compliance experts were, interestingly, scoring most of the uses, including the overlooked ones, as `very likely' ($>$35\% of uses) to be adopted. In this case, a chi-squared test results confirmed that the distributions of scores for adoption potential significantly differed between the two cohorts, with compliance experts generally giving higher scores.   

\vspace{0.3em}
\noindent
\textbf{V. Transformational impact.}
Developers were slightly more conservative in estimating the potential for transformative impact of the uses (Figure \ref{fig:eval}), assigning the largest proportion of uses a `neutral' score ($\sim$20\%). In contrast, compliance experts gave the highest proportion of `very likely' scores ($>$25\% of the uses) for both all and overlooked ones.
Similarly as for the adoption potential scores, a chi-squared test results confirmed that the distributions of scores for transformational potential significantly differed between the two cohorts.

\vspace{0.3em}
\noindent
\textbf{VI. (Perceived) riskiness.}
 Each use was rated by three compliance experts. To obtain the ground truth label, we required that at least 2 of the 3 labels were aligned. By comparing these ground truth labels with the \emph{RiskLabelling} labels, we found that 94.5\% of the uses were correctly classified, with an almost perfect Cohen's Kappa agreement of 92.2\%. However, the inter-rater agreement among the three annotators was only moderate, with a Fleiss' Kappa score of 49.1\%, suggesting the task is challenging and that participants might have defaulted to the provided labels.

For example, participants disagreed with the LLM's limited or low-risk classification for uses such as \emph{verifying the identity of customers during transactions by banks, use \#19}, and \emph{identifying obstacles and people to avoid collisions by robots, use \#56}. For \emph{use \#19}, they commented that it should be high risk due to the \emph{“high chance for fraud”} and the possibility that the \emph{“AI system could see the PIN of the bank card!”}. For \emph{use \#56}, two annotators voted for a high-risk label because \emph{“in the case of misuse or malfunctioning, the AI could lead to serious harm for individuals”} and \emph{“[...] put human lives at risk.”}

On the other hand, the participants did not agree with the high-risk classification for \emph{assisting law enforcement agencies in criminal investigations by identifying suspects in video footage, use \#85}. Two of them thought this use should be classified as \emph{prohibited} in the EU, as it could lead to violations of privacy rights. The LLM did not classify it as such because the identification from footage is not in real-time, which is a requirement for prohibited uses specified in Article 5(1)(d). The third annotator, however, suggested downgrading the risk classification to limited or low risk because the use is \emph{“necessary to provide proof and existence of criminal activities and facilitate law enforcement work”}. These examples demonstrate the subtleties in the risk assessment task, including the interpretation of the use context and the annotators' personal viewpoints \cite{hupont2022landscape}, which partly explain the lower inter-rater agreement among our participants.

As shown in Figure \ref{fig:eval}, developers thought that most of the uses are only `slightly' to `moderately risky' for society (approximately 20-25\%), and not at all risky for the environment (approximately 50-60\%). This contrasts with our risk classification finding that over 50\% of the uses are either high-risk or prohibited according to the EU AI Act. These results highlight the challenge developers face in identifying and classifying the riskiness of AI uses.

\vspace{0.3em}
\noindent
\textbf{VII. Usefulness.}
Finally, we studied the extent to which the two cohorts of AI practitioners found our framework useful in assisting with tasks such as envisioning AI uses and assessing associated risks.

AI compliance experts found \emph{ExploreGen}'s output particularly useful. For example, L01 mentioned that a tool \emph{“classifying [uses] in different ways and offering various uses of those [technologies], would be very useful in my job, [...] because it would help me look at things in a different way.”} L03 stated, \emph{“I enjoyed it [...] I think it’s really helpful to kind of envision what will be the future use of AI and then think about how it will impact society and the environment. I think it’s a good exercise for someone working in the tech space in general,”} and \emph{“... it will also be useful for people who want to understand the technology, like people impacted by the technology and the public.”}
One participant from a major tech company developing FRA technologies expressed excitement upon discovering uses they are currently working on, particularly in risk and compliance assessment. They also found inspiration for new potential use cases, stating, \emph{“We are putting more effort into going into the [domain X], and that could be a good use.”}
L04 was particularly engaged with the risk-classification output provided by our tool. For instance, they focused on the use \emph{identifying personnel by logistics companies to improve the efficiency of cargo handling, use \#104,} and agreed with the low-risk classification. They noted that \emph{“[A major company] has just gotten a judgment in its favor that very far-reaching analytics in its plants in [country Y] are permissible.”}.
L03 was also inspired to think about the risks of the presented uses. They deliberated about the use \emph{verifying patient identity in medical settings, use \#10}, which is classified as low risk, but they thought it could incur many risks as \emph{“services like this [...] can be exclusionary to certain, especially marginalized communities.”} They concluded, \emph{“I would look into developing this, but I would consider this a high-risk use depending on the context and on the decision that’s being made by verifying.”}

AI developers, on the contrary, initially struggled to identify the application of our tool in their everyday work. While interested in exploring the presented uses, they frequently asked for more details and insights on specific uses. For instance, D02 expressed feeling overwhelmed by the comprehensive list of uses: \emph{“I imagine [I am] developing that, and put a lot of cognitive load in each case and then imagining how it will work and how it will be developed.”}.
During the interviews, it became clear that developers, especially those working on business products, have less opportunity to use a tool like \emph{ExploreGen} because they typically do not engage in extensive brainstorming and reflexivity. Instead, they usually receive well-defined uses to develop. For example, D03 commented, \emph{“I've been working with products and generally you start with a use that you want to develop [...] and then you work backwards and maybe a technology is not useful for that particular problem.”} D03 also stated, \emph{“For most of the people I speak with, it seems like more of an afterthought than like an active design. [You think] what could be the risks kind of post hoc?”} They added, \emph{“But I think people are generally getting a little bit better at that now because I think people are seeing that AI is progressing quite fast...”} 
For these reasons, developers appreciated the color-coding of the use risk levels, as it provided a quick overview of the more or less risky domains, contexts, and uses. One participant noted surprise at seeing a similar use having different risk levels in two domains, finding the tool helpful for educating them about the EU AI Act and its domain-based risk classification.
D01, who holds the most senior role among the developer participants, stated: \emph{“[We] have a brainstorming session on first of all, understanding if AI is really needed to solve the problem or not[...]”} They added about our tool: \emph{“It will be very helpful for me or someone in my team to get a first sense of the risks involved...”}
Generally, developers preferred the second task in the study, where they could focus on a subset of uses and scrutinize them in detail, as this aligns more closely with their job responsibilities. Additionally, those in senior roles and closer to R\&D found our tool more useful for brainstorming and deliberation tasks compared to junior developers and those working in business production.

Both AI developers and AI compliance experts agreed that a tool like ours would be most useful during the \emph{design} stage of AI development. Moreover, several participants indicated they would use it throughout \emph{all} stages, as noted by L4: \emph{“I don't think one stage is more important than the other. I think there are different risks at different stages.”}

\subsection{Discussion}
\label{sec_discussion}
The findings from nine user studies revealed the potential of our proposed LLM framework \emph{ExploreGen} to facilitate reflectivity, ideation, and deliberation for both AI developers and compliance experts—tasks that are increasingly essential but often challenging to perform \cite{liang2024s,prunkl2021institutionalizing}. Our tool contributes to the existing body of research calling for \cite{sherman2024ai} and exploring \cite{buccinca2023aha,wang2024farsight} LLMs as a means to support responsible AI design. 

\subsection{Implications}

\noindent
\textbf{Brainstorming in AI Developer Teams.} 
\emph{ExploreGen} successfully generated realistic uses that practitioners were not very familiar with, many of which were rated as having high adoption potential and transformational impact. Developers found the overview of uses contextualized across various domains, along with their risk levels, to be informative. Some saw the tool's value during brainstorming meetings while deliberating on which directions for technology applications to pursue. Additionally, they expressed interest in a tool with a more in-depth analysis of specific uses, allowing to break down the associated risks of the use they are developing and be informed about similar risks faced by different uses.

\noindent \textbf{Bridging Risk Perception with Compliance.} 
Compliance experts agreed with the risk classifications provided by \emph{RiskLabelling}, though they noted that subtle changes in the context of use might alter the classification level. Despite more than 50\% of the FRA technology uses being classified as high risk or prohibited, practitioners perceived them as mostly only slightly risky for society and not at all for the environment. However, due to the size of the datasets and computational demands, energy consumption is becoming an important consideration for FRA technology \cite{hassel2023governing}, highlighting a disconnect in AI practitioners' understanding of all the technology's impacts.
 
\noindent \textbf{Data-driven Deliberation for Compliance Experts.} 
Compliance experts saw more direct applications of \emph{ExploreGen} in its current form for their work, as they often explore various (often unintended or unexpected) contexts of use for a given technology. They found the tool very helpful for this task. They also appreciated the breakdown of uses across various domains and risk levels and wanted features allowing for additional breakdowns (e.g., according to the subjects or types of risk).

\subsection{Limitations and Future Work}
\noindent
\textbf{{LLM Method Shortcomings.}}
The use of LLMs presents four main challenges. First, the generated uses, and risks may be limited to the training set and biased \cite{luccioni2024stable}, potentially overlooking important aspects. Enhancements could include fine-tuning \cite{hu2023llm} or augmenting with specialized datasets (e.g., from AI Incident Database \cite{mcgregor2021preventing}). Second, there is a risk of incorrect outputs due to LLM hallucinations \cite{mittelstadt2023protect}. \emph{UsesGen} identified 6\% unrealistic uses, which were removed. Future research could explore combining classifiers and manual checks to ensure accuracy \cite{mittelstadt2023protect}. Third, LLMs may be overly conservative, missing risky edge-case uses due to built-in guardrails. Last, presenting LLM outputs to users could create a false sense of security \cite{pataranutaporn2023influencing}. Ongoing research in human-AI interaction offers strategies to mitigate these issues, such as designing cognitive forcing functions \cite{buccinca2021trust} and skill improvement \cite{buccinca2024towards}.

\noindent
\textbf{Difficulty of Risk Classification.} 
We focused on labeling prohibited and high-risk uses, with the remainder classified as limited or low risk. However, the EU AI Act includes an additional classification label, transparency risk, which we omitted due to the task's inherent complexity arising from ambiguities in the Act's wording \cite{veale2021demystifying}. These ambiguities, resulting from the interplay between technical and legal jargon, pose challenges even for professionals in the field, as reflected in the moderate inter-rater agreement among our user study participants. Additionally, while the five risk categories aid in classification, practical variations in each use ultimately determine their final classification.

\noindent \textbf{Generalizability.} While we evaluated our framework with 25 AI practitioners on the case of FRA technology, future work should explore its applicability to other technologies and involve a larger set of AI practitioners, researchers, and the general public.

\bibliography{aaai24}

\clearpage
\appendix
\onecolumn

\section{Appendix}
\subsection{(A) UsesGen}

\begin{figure*}[htbp!]
\centering
\includegraphics[width=0.777\textwidth]{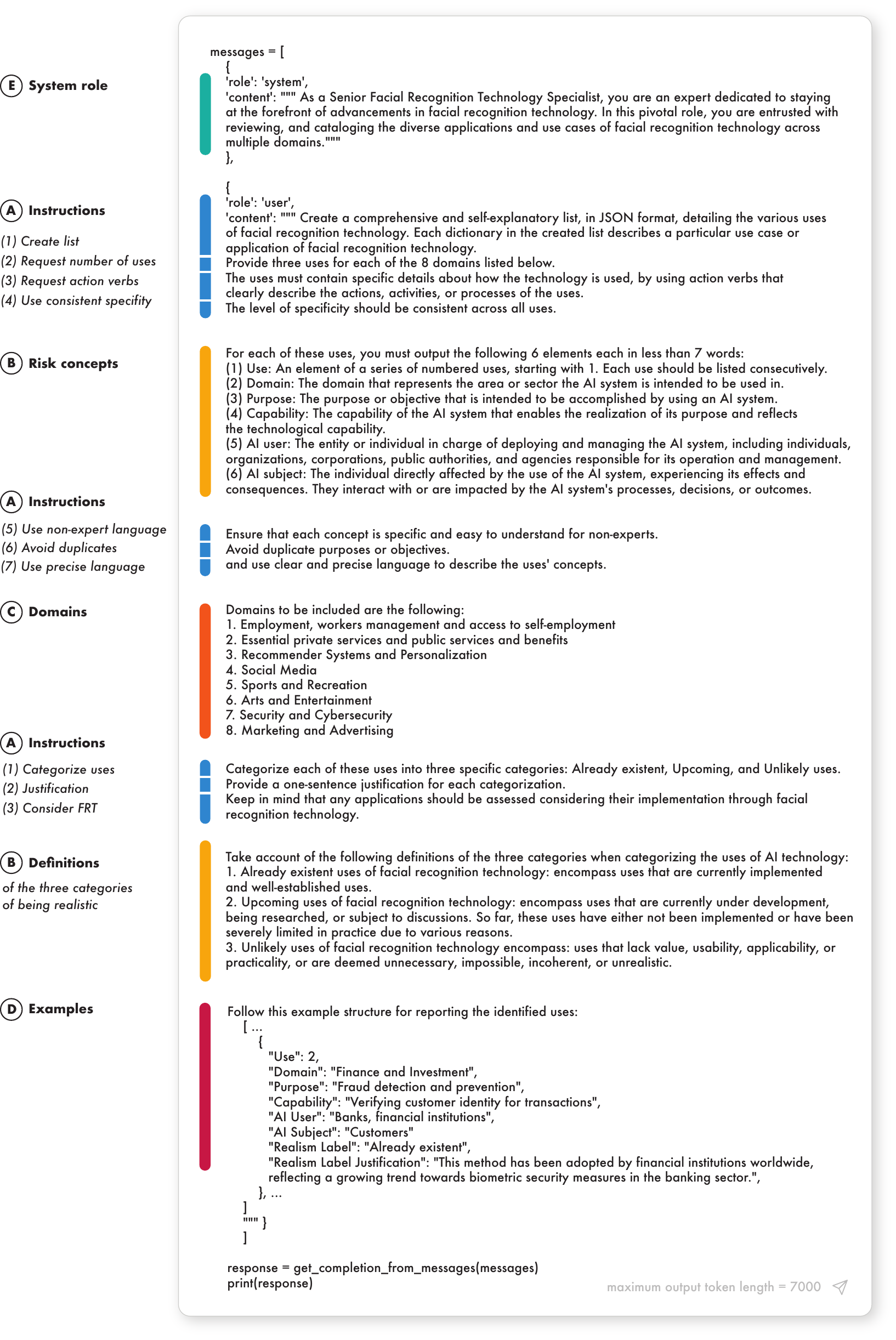}
\caption{\textbf{UsesGen.} The prompt generates a list of uses for a given AI technology, e.g., FRA. These LLM-generated uses are required to be outputted in the format of 5 risk concepts (domain, purpose, capability, AI user, AI subject) \cite{Golpayegani2023Risk}. This format allows the subsequent RiskLabelling prompt to evaluate the risk of a given AI technology use.To identify the most comprehensive and realistic list of LLM-generated uses, we examined different UsesGen configurations. These prompt configurations included the model temperature, number of requested uses per domain (2 or 3), and prompt elements (Variation 1-3). Variation 1 of UsesGen encompassed an instruction (A), definitions of risk concepts and the three categories of being realistic (B), and domains (C), that correspond to the necessary elements (Figure \ref{Fig:configurations}). In Variation 2, we introduced the system role (D), while in Variation 3, we included an additional five examples (E).} 
\label{Fig:Generation}
\end{figure*}


\begin{table}[h!]
\caption{We identified a list of 46 domains from the EU AI Act \cite{EUACT2023} and an interactive session with our research team ($N$=8). Among these, 40 domains are from the EU AI Act, and 6 additional domains -- not covered by the previous list of 40 domains -- were identified during the session with our team (indicated by an asterisk *).}
\small
\label{app:domains}
\resizebox{\textwidth}{!}{%
\begin{tabular}{clcl}
\toprule
\textbf{No.} & \textbf{Domain} & \textbf{No.} & \textbf{Domain} \\ \midrule
\textbf{1} & Biometric identification and categorization of natural persons & \textbf{24} & Democracy \\
\textbf{2} & Family & \textbf{25} & Media and Communication \\
\textbf{3} & Romantic relationships and friendships & \textbf{26} & Accessibility and Inclusion \\
\textbf{4} & Health and Healthcare & \textbf{27} & Energy \\
\textbf{5} & Well-being & \textbf{28} & Military and Defense \\
\textbf{6} & Human-Computer Interaction & \textbf{29} & Administration of justice and democratic processes \\
\textbf{7} & Finance and Investment & \textbf{30} & Government Services and Administration \\
\textbf{8} & Education and vocational training & \textbf{31} & Diplomacy and Foreign Policy \\
\textbf{9} & Employment, workers management and access to self-employment & \textbf{32} & Food Safety and Regulation \\
\textbf{10} & Essential private services and public services and benefits & \textbf{33} & Crisis Management and Emergency Response \\
\textbf{11} & Recommender Systems and Personalization & \textbf{34} & Humanitarian Aid \\
\textbf{12} & Social Media & \textbf{35} & Transport and Logistics \\
\textbf{13} & Sports and Recreation & \textbf{36} & Urban Planning \\
\textbf{14} & Arts and Entertainment & \textbf{37} & Counterterrorism \\
\textbf{15} & Security and Cybersecurity & \textbf{38} & Environment and Sustainability \\
\textbf{16} & Marketing and Advertising & \textbf{39} & International Law Enforcement and Cooperation \\
\textbf{17} & Agriculture and Farming & \textbf{40} & Climate Change Mitigation and Adaptation \\
\textbf{18} & Entrepreneurship & \textbf{41} & Gaming and interactive experiences* \\
\textbf{19} & Autonomous Robots and Robotics & \textbf{42} & Hobbies* \\
\textbf{20} & Innovation and Research & \textbf{43} & Smart home* \\
\textbf{21} & Management and Operation of critical infrastructure & \textbf{44} & Social and Community Services* \\
\textbf{22} & Law enforcement & \textbf{45} & Public and private transportation* \\
\textbf{23} & Migration, Asylum and Border control management & \textbf{46} & Interpersonal Communication* \\ \bottomrule
\end{tabular}%
}
\end{table}

\subsection{(B) GT: FRA Uses from the Literature}

%
1. Access control for buildings, areas, storage places, banks' vaults and lockers.   \cite{7935088, 8281895, 6718253, 9137927, 7377975, 7280401, 8746942, 9339602, 9240416, 9721723, 9639574, 9687215, 9166682, 9609705, 9912953, 10169208, 10071654, 10024248, 9544851, 9795021, 9648153, 9761288, 9825108, 9803092, 10080678, 8947836, 10037567, 9236972, 9133968, 9931155, 10.1145/3038924, 10.1145/3507623.3507627} \\
2. Access control for technology, secure networks, and resources.   \cite{9964836, 9912953, 10.1145/3038924} \\
3. Detect unauthorized personnel.   \cite{9236972} \\
4. Securely register and manage visitors.   \cite{9236972, 10157843} \\
5. Provide real-time information about visitors in high-profile buildings.   \cite{10146655} \\
6. Identify men in female-reserved coaches or women-only areas.   \cite{9027290} \\
7. Visualize building occupancy and peak hours and determine foot traffic patterns.   \cite{9952691} \\
8. Identify individuals approaching homes for example through smart doorbells.   \cite{9027290, 9639441} \\
9. Trigger an alarm when unidentified individuals enter a home.   \cite{8746942} \\
10. Enable unlocking devices and apps.   \cite{7935088, 8281895, 6718253, 9596066, 8746942, 9964836, 9639574, 10165907, 10024248, 9833871, 9639441, 7029643, 8409387, 7126344, 9206794, 8592634, 9515221, 9236972, 10074417} \\
11. Enable unlocking cars and driving them.   \cite{8281895, 7414809, 9761288} \\
12. Notify owners if someone attempts to steal their car.   \cite{8281895} \\
13. Track worker and student attendance.   \cite{7935088, 8281895, 8825049, 9166595, 9829079, 8987961, 9596066, 9590122, 9964836, 9339602, 9721723, 9493844, 9166682, 10165907, 9609705, 10071654, 9998001, 9544851, 9795021, 9648153, 9515221, 9952691, 9310090, 9236972, 10157843, 10054833, 9924775} \\
14. Check if students and workers comply with regulations.   \cite{9964836, 10157843} \\
15. Monitor and track students' activities and performance to aid university examinations.   \cite{10165907, 9743134, 10157843} \\
16. Identify demographic attributes of individuals, including gender, age, ethnicity, and sexual orientation.   \cite{8756700, 9137927, 9027290, 9964836, 9210982, 10.1145/3375627.3375820} \\
17. Perform profiling to identify patterns and characteristics of individuals or groups.   \cite{9687215} \\
18. Implement targeted recognition systems that customize responses or services based on a person's ethnicity or gender.   \cite{9639441} \\
19. Customize advertisements and promotions for targeted marketing.   \cite{9027290} \\
20. Identify individuals' shopping habits for personalized discounts and offerings.   \cite{9639441} \\
21. Evaluate consumer satisfaction.   \cite{9639441, 10.1145/3375627.3375820} \\
22. Monitor wait times and moods in check-out or customer service lines.   \cite{9639441} \\
23. Personalized recommendations, surroundings, and services for smart homes, automotive environments, and travel industry.   \cite{9027290, 7414809} \\
24. Identify and track criminals, suspects, stalkers, or terrorists.   \cite{6718253, 9137927, 9596066, 9027290, 9166682, 10165907, 9609705, 9544851, 9639441, 9106661, 9803092, 10080678, 9236972, 10.1145/3194452.3194479, 10.1145/3375627.3375820, 10.1145/3561877.3561888} \\
25. Recording an arrested individual's facial information.   \cite{9310090, 8987345} \\
26. Gathering, analyzing, and interpreting evidence from a crime scene or incident.   \cite{9027290, 8877744, 9310090, 10.1145/3293353.3293432} \\
27. Register traffic violations by rental transport users.   \cite{9027290} \\
28. Establish a unified penalty system, collecting fines for various violations such as fare dodging.   \cite{9027290} \\
29. Define groups of people and create whitelists for VIPs and blacklists for unwanted individuals.   \cite{10165907} \\
30. Compare individuals against watchlists containing names, identifiers, or attributes of known individuals of interest or potential risks.   \cite{9137927, 9206794, 9761288, 10074417} \\
31. Prevent child exploitation and abduction.   \cite{9025674} \\
32. Search for and identify missing persons.   \cite{9596066, 9027290, 9632692, 9544851, 9639441, 10074417, 10.1145/3507623.3507627} \\
33. Identify abusive law enforcement officers.   \cite{9639441} \\
34. Implement face tagging in images.   \cite{9137927, 7280401, 9596066, 10165907, 5981788, 9639441, 9106661, 10074417, 10109304, 10.1145/3485447.3512212} \\
35. Create digital photo books.   \cite{9025674} \\
36. Alert individuals when photographs with their faces are posted online.   \cite{8746942} \\
37. Detect and report inappropriate pictures using facial recognition and pattern analysis.   \cite{9632692} \\
38. Group photos based on individuals present.   \cite{10.1145/3507623.3507627} \\
39. Conduct face scans to search for specific individuals in pictures.   \cite{8746942, 10.1145/2983402.2983420, 10.1145/3375627.3375820} \\
40. Use selfies to find users' doppelganger in a database of recognized paintings.   \cite{9639441} \\
41. Recognize when fake profiles use someone else's face.   \cite{9639441} \\
42. Offer friend suggestions on social media platforms.   \cite{9632692} \\
43. Prevent online dating fraud.   \cite{9632692} \\
44. Enable face authentication-based mobile payments and other banking services.   \cite{9027290, 8746942, 9964836, 9721723, 9639574, 9687215, 9639441, 9648153, 7126344, 9761288, 9833501} \\
45. Compare a customer's face during ATM usage with database to reduce fraudulent activities.   \cite{6718253, 10165907} \\
46. Payments on public transport (E.g. metro trains, public buses, toll gates).   \cite{9027290} \\
47. Identify patients and facilitate check-in and other processes for patients, for example, provide notifications to respective doctors, generate e-prescriptions for patients, update recent improvements in a patient's case study.   \cite{9027290, 9250002} \\
48. Identify healthcare staff.   \cite{9027290, 9833871} \\
49. Track staff and patients to keep a record of the movement and presence of both staff members and patients within a healthcare facility.   \cite{9027290, 9250002} \\
50. Streamline and manage medicine distribution in healthcare settings.   \cite{9027290, 9639441} \\
51. Diagnose or support detection of diseases in individuals.   \cite{9027290, 9515221, 10100517} \\
52. Retrieve critical information of people in emergencies, such as their blood group.   \cite{9027290} \\
53. Conduct real-time mental health tests.   \cite{10.1145/3495018.3501073} \\
54. Provide automatic floor selection in elevators for elderly and individuals with disabilities.   \cite{9027290} \\
55. Assist individuals with impairments by identifying their friends and people, providing reminders of names, and relevant information about them.   \cite{9206794, 9952691, 10.1145/3173574.3173789, 10.1145/3194452.3194479} \\
56. Assist individuals with visual impairments in taking photos.   \cite{10.1145/3173574.3173789} \\
57. Initiate robot operations only when it recognizes an operator in its workspace.   \cite{9687215} \\
58. Enable assistive robots to recognize individuals in a home environment.   \cite{10.1145/3378184.3378225} \\
59. Facilitate voter identification processes.   \cite{10165907, 10134227} \\
60. Provide tailored learning experiences to suit individual needs.   \cite{8333315} \\
61. Create multimedia content using facial recognition.   \cite{5630848, 8877744, 9123123, 9106661} \\
62. Identify and differentiate between various characters and actors in movies, making it highly beneficial for content discovery and delivery platforms that seek to provide content based on specific characters or actors.   \cite{7881542} \\
63. Overlay cosmetic changes for users.   \cite{9639441} \\
64. Use photo and video filters and special beautification effects.   \cite{9639441, 10.1145/3313129} \\
65. Streamline check-ins and boarding and reduce waiting times at airports.   \cite{6718253, 9137927, 7280401, 9027290, 9639441, 9761288, 9803092} \\
66. Identity travelers at border crossings to automate border crossing procedures (traveler identification, biometric passport, passport-checking).   \cite{7935088, 8965805, 6718253, 9027290, 9339602, 9639574, 9210982, 9632692, 9025674, 8545076, 10074417, 8987345, 10.1145/3038924} \\
67. Verify documents, such as passports, visas, and driver's licenses.  \cite{7280401, 10165907, 10074417, 10.1145/3507623.3507627} \\
68. Identify and verify people in train stations and stadiums.  \cite{9803092} \\
69. Detect fatigue in individuals.  \cite{7814911} \\
70. Detect emotions in individuals.  \cite{10080678, 10.1145/3375627.3375820} \\
71. Facilitate recruitment processes by informing hiring decisions and help job interviewers to view candidate’s previous records.  \cite{10.1145/3194452.3194479, 10.1145/3375627.3375820} \\
72. Help people recognize faces by using smart glasses to display names and social network activities of identified people.  \cite{10.1145/1959826.1959857, 10.1145/3194452.3194479} \\
73. Monitor and surveil people.  \cite{8756700, 8965805, 9137927, 7377975, 9027290, 9065100, 9339602, 9240416, 5630848, 9687215, 9166682, 8600370, 8756593, 9206794, 10080678, 9515221, 8947836, 9310090, 9133968, 10.1145/3375627.3375820} \\
74. Facilitate tourists in meeting new people.  \cite{10.1145/3194452.3194479} \\

\subsection{(C) RiskLabelling}

\begin{figure*}[htbp!]
\centering
\includegraphics[width=0.777\textwidth]{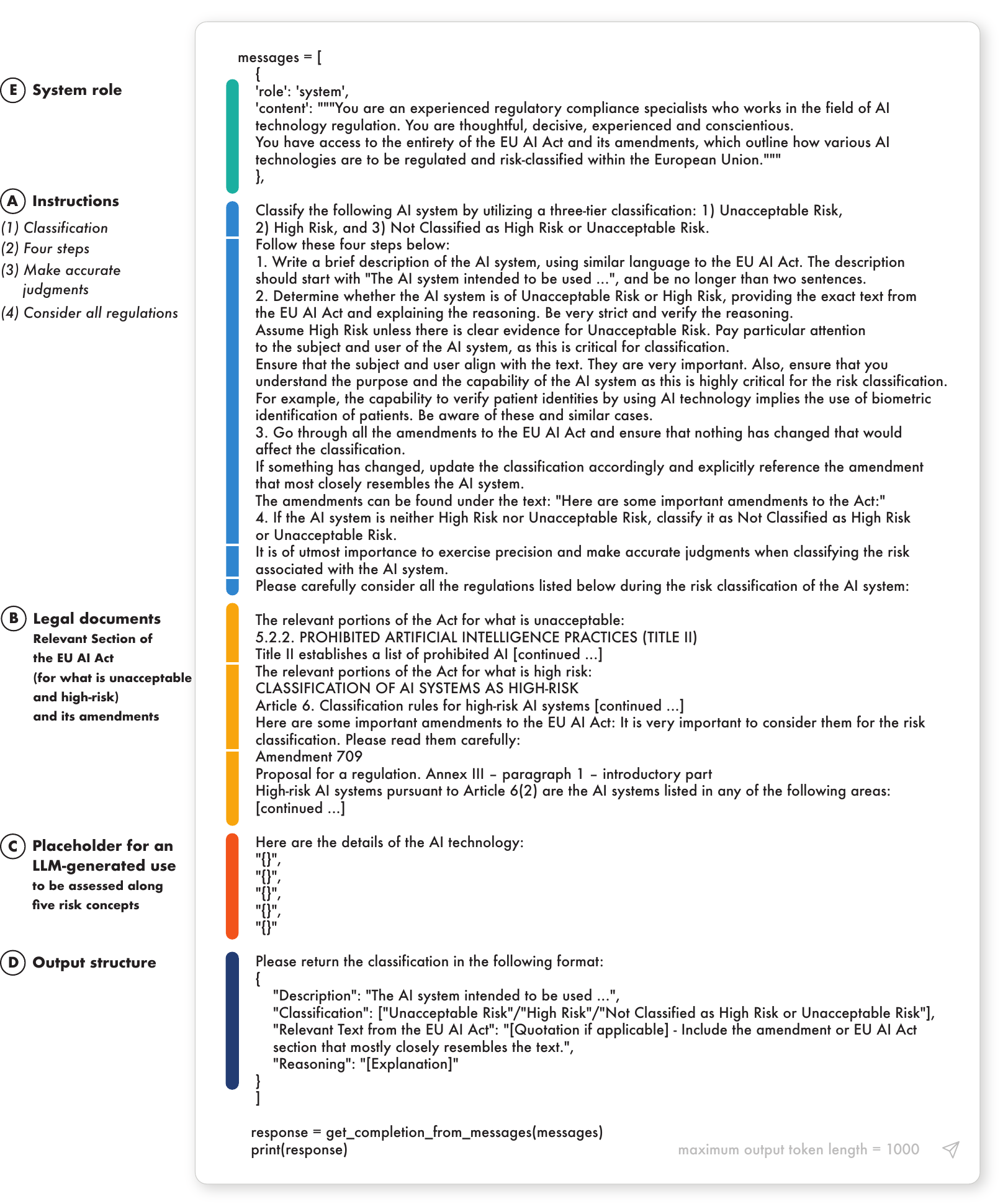}
\caption{\textbf{RiskLabelling}. The prompt evaluates how risky the LLM-generated uses are. Specifically, the objective is to classify the LLM-generated uses of the list into unacceptable risk, high risk, or neither unacceptable nor high risk.
The Risk Assessment includes Instructions (A), Relevant Sections of the EU AI Act for what is unacceptable, high risk, and the amendments (i.e., Annex III and its amendments) (B), an LLM-generated use (C), Output Structure (D), and a System Role (E).
}
\label{Fig:Risk}
\end{figure*}

\begin{table*}[!htb]
\caption{Examples of prohibited (P), high-risk (HR), and limited or low-risk (LR) LLM-generated uses along with the reasoning for use classification based on the EU AI Act provided by GPT-4, as part of our ExploreGen framework.}
\label{tab:riskex}
\small
\centering
\begin{tabular}{p{0.03\textwidth} p{0.4\textwidth} p{0.45\textwidth}}
\toprule
\textbf{Class} & \textbf{LLM-generated Use} & \textbf{Reasoning for Use Classification} \\ \midrule
\textbf{P} & \textbf{1)} \textbf{Domain}: Security and Cybersecurity,
\textbf{Purpose}: Surveillance,
\textbf{Capability}: Identifying individuals in surveillance footage,
\textbf{AI User}: Law Enforcement, Security Companies,
\textbf{AI Subject}: General Public
& \textbf{1)} Prohibited due to the use of real-time remote biometric identification in publicly accessible spaces for law enforcement, which falls under the EU AI Act Article 5(1)(d). \\ 
\hdashline
\textbf{HR}  & \textbf{2)} \textbf{Domain}: Smart home,
\textbf{Purpose}: Monitoring child safety,
\textbf{Capability}: Alerting when unrecognized faces are detected,
\textbf{AI User}: Parents, security companies,
\textbf{AI Subject}: Children
& \textbf{2)} High Risk due to the use of biometric identification, which falls under the EU AI Act Article 6(2) and Annex III, Section 1(a).  \\
\hdashline
\textbf{LR} & \textbf{3)} \textbf{Domain}: Gaming and interactive experiences,
\textbf{Purpose}: Enhancing player immersion,
\textbf{Capability}: Translating player's facial expressions into game,
\textbf{AI user}: Game developers, VR platforms,
\textbf{AI subject}: Gamers
& \textbf{3)} Limited or Low Risk due to its application in gaming for enhancing immersion without significant risk to fundamental rights or safety. \\ 
\bottomrule
\end{tabular}%
\end{table*}

\subsection{(D) List of FRA Uses Generated by UsesGen}
\label{app:uses} 

\begin{longtable}{p{0.3\textwidth}p{0.3\textwidth}p{0.3\textwidth}}
\caption{LLM-Generated List of FRA Uses created with UsesGen. Each of the uses comes with 6 elements which are the Use ID (e.g., 1), Domains (e.g., Biometric identification and categorisation of natural persons), Purpose (e.g., Secure access control), Capability (e.g., Verifying identity through facial features), AI User (e.g., Security firms, corporations), AI Subject (e.g., Employees, visitors).}
\label{tab:llmuses}\\
\toprule
\multicolumn{3}{c}{\textbf{LLM-Generated Uses FRA}} \\ \midrule
\endfirsthead
\multicolumn{3}{c}%
{{\bfseries Table \thetable\ continued from previous page}} \\
\toprule
\multicolumn{3}{c}{\textbf{LLM-Generated Uses of a given AI Technology (FRA)}} \\ \midrule
\endhead
\hline
\endfoot
\endlastfoot
Use: 1, & Use: 47, & Use: 93, \\
Domain: Biometric identification and categorisation of natural persons, & Domain: Marketing and Advertising, & Domain: Diplomacy and Foreign Policy, \\
Purpose: Secure access control, & Purpose: Customer behaviour analysis, & Purpose: Improving international relations, \\
Capability: Verifying identity through facial features, & Capability: Analysing customer reactions to ads, & Capability: Identifying foreign officials in meetings, \\
AI User: Security firms, corporations, & AI User: Advertisers, Marketing Agencies, & AI User: Diplomats, foreign affairs departments, \\
AI Subject: Employees, visitors & AI Subject: Consumers & AI Subject: Foreign officials \\ 
{\dotfill} & {\dotfill} & {\dotfill} \\
Use: 2, & Use: 48, & Use: 94, \\
Domain: Biometric identification and categorisation of natural persons, & Domain: Marketing and Advertising, & Domain: Food Safety and Regulation, \\
Purpose: Criminal identification, & Purpose: Personalised marketing, & Purpose: Ensuring food safety compliance, \\
Capability: Matching faces to criminal databases, & Capability: Recognising customer preferences for personalised marketing, & Capability: Identifying individuals in food production facilities, \\
AI User: Law enforcement agencies, & AI User: Retailers, E-commerce Platforms, & AI User: Food safety regulators, food companies, \\
AI Subject: Suspects, criminals & AI Subject: Customers & AI Subject: Food production workers \\
{\dotfill} & {\dotfill} & {\dotfill} \\
Use: 3, & Use: 49, & Use: 95, \\
Domain: Biometric identification and categorisation of natural persons, & Domain: Agriculture and Farming, & Domain: Food Safety and Regulation, \\
Purpose: Personalised advertising, & Purpose: Livestock monitoring and management, & Purpose: Improving food traceability, \\
Capability: Identifying demographic characteristics from faces, & Capability: Identifying individual animals and tracking their health, & Capability: Identifying individuals handling food products, \\
AI User: Advertisers, marketers, & AI User: Farmers, livestock managers, & AI User: Food companies, regulators, \\
AI Subject: Consumers & AI Subject: Livestock & AI Subject: Food handlers, consumers \\
{\dotfill} & {\dotfill} & {\dotfill} \\
Use: 4, & Use: 50, & Use: 96, \\
Domain: Family, & Domain: Agriculture and Farming, & Domain: Food Safety and Regulation, \\
Purpose: Family photo organisation, & Purpose: Crop health assessment, & Purpose: Enhancing food quality control, \\
Capability: Sorting photos based on facial recognition, & Capability: Detecting plant diseases and pest infestations, & Capability: Monitoring and identifying individuals in quality control, \\
AI User: Family members, photo storage platforms, & AI User: Farmers, agronomists, & AI User: Food companies, quality control agencies, \\
AI Subject: Family members & AI Subject: Crops & AI Subject: Quality control personnel \\
{\dotfill} & {\dotfill} & {\dotfill} \\
Use: 5, & Use: 51, & Use: 97, \\
Domain: Family, & Domain: Agriculture and Farming, & Domain: Crisis Management and Emergency Response, \\
Purpose: Child safety monitoring, & Purpose: Harvest optimisation, & Purpose: Identifying victims in disaster zones, \\
Capability: Identifying unfamiliar faces in child's vicinity, & Capability: Determining optimal harvest times based on crop maturity, & Capability: Scanning and matching faces in real-time, \\
AI User: Parents, child safety apps, & AI User: Farmers, agricultural consultants, & AI User: Emergency response teams, \\
AI Subject: Children & AI Subject: Crops & AI Subject: Disaster victims \\
{\dotfill} & {\dotfill} & {\dotfill} \\
Use: 6, & Use: 52, & Use: 98, \\
Domain: Family, & Domain: Entrepreneurship, & Domain: Crisis Management and Emergency Response, \\
Purpose: Elderly care assistance, & Purpose: Customer identification and penalisation, & Purpose: Verifying identity of emergency responders, \\
Capability: Recognising signs of distress or confusion, & Capability: Recognising customers and tailoring services to their preferences, & Capability: Authenticating faces against a database, \\
AI User: Caregivers, elderly care facilities, & AI User: Business owners, customer service representatives, & AI User: Emergency response agencies, \\
AI Subject: Elderly individuals & AI Subject: Customers & AI Subject: Emergency responders \\
{\dotfill} & {\dotfill} & {\dotfill} \\
Use: 7, & Use: 53, & Use: 99, \\
Domain: Romantic relationships and friendships, & Domain: Entrepreneurship, & Domain: Crisis Management and Emergency Response, \\
Purpose: Social media tagging, & Purpose: Security enhancement, & Purpose: Locating missing persons, \\
Capability: Identifying friends in photos for tagging, & Capability: Verifying identities to prevent unauthorised access, & Capability: Comparing faces in crowds to missing persons, \\
AI User: Social media platforms, users, & AI User: Business owners, security personnel, & AI User: Search and rescue teams, \\
AI Subject: Social media users & AI Subject: Employees, customers & AI Subject: Missing persons \\
{\dotfill} & {\dotfill} & {\dotfill} \\
Use: 8, & Use: 54, & Use: 100, \\
Domain: Romantic relationships and friendships, & Domain: Entrepreneurship, & Domain: Humanitarian Aid, \\
Purpose: Dating app matching, & Purpose: Employee attendance tracking, & Purpose: Distributing aid to verified recipients, \\
Capability: Matching faces to user preferences, & Capability: Monitoring employee check-ins and check-outs, & Capability: Recognising faces to confirm identity, \\
AI User: Dating apps, users, & AI User: Business owners, HR managers, & AI User: Aid organisations, \\
AI Subject: Dating app users & AI Subject: Employees & AI Subject: Aid recipients \\
{\dotfill} & {\dotfill} & {\dotfill} \\
Use: 9, & Use: 55, & Use: 101, \\
Domain: Romantic relationships and friendships, & Domain: Autonomous Robots and Robotics, & Domain: Humanitarian Aid, \\
Purpose: Friendship analysis, & Purpose: Human-robot interaction, & Purpose: Reuniting separated families, \\
Capability: Analysing interaction patterns in photos, & Capability: Recognising and responding to human faces and expressions, & Capability: Matching faces to find family members, \\
AI User: Social media platforms, users, & AI User: Robot developers, operators, & AI User: Refugee agencies, \\
AI Subject: Social media users & AI Subject: Robot users & AI Subject: Separated family members \\
{\dotfill} & {\dotfill} & {\dotfill} \\
Use: 10, & Use: 56, & Use: 102, \\
Domain: Health and Healthcare, & Domain: Autonomous Robots and Robotics, & Domain: Humanitarian Aid, \\
Purpose: Patient identification, & Purpose: Robot navigation, & Purpose: Tracking disease spread in refugee camps, \\
Capability: Verifying patient identity in medical settings, & Capability: Identifying obstacles and people to avoid collisions, & Capability: Identifying individuals in contact with infected persons, \\
AI User: Hospitals, clinics, & AI User: Robot developers, operators, & AI User: Health organisations, \\
AI Subject: Patients & AI Subject: People in robot's environment & AI Subject: Refugees \\
{\dotfill} & {\dotfill} & {\dotfill} \\
Use: 11, & Use: 57, & Use: 103, \\
Domain: Health and Healthcare, & Domain: Autonomous Robots and Robotics, & Domain: Transport and Logistics, \\
Purpose: Disease diagnosis, & Purpose: Personalised robot services, & Purpose: Enhancing security at transport hubs, \\
Capability: Identifying disease symptoms on faces, & Capability: Recognising specific individuals for personalised interactions, & Capability: Detecting and alerting on known criminals, \\
AI User: Healthcare professionals, AI diagnostic tools, & AI User: Robot developers, operators, & AI User: Transport authorities, \\
AI Subject: Patients & AI Subject: Robot users & AI Subject: Travellers \\
{\dotfill} & {\dotfill} & {\dotfill} \\
Use: 12, & Use: 58, & Use: 104, \\
Domain: Health and Healthcare, & Domain: Innovation and Research, & Domain: Transport and Logistics, \\
Purpose: Mental health assessment, & Purpose: Participant identification in research studies, & Purpose: Improving efficiency in cargo handling, \\
Capability: Analysing facial expressions for emotional state, & Capability: Recognising and tracking participants in studies, & Capability: Identifying authorised personnel for cargo access, \\
AI User: Psychologists, mental health apps, & AI User: Researchers, scientists, & AI User: Logistics companies, \\
AI Subject: Patients & AI Subject: Research participants & AI Subject: Cargo handlers \\
{\dotfill} & {\dotfill} & {\dotfill} \\
Use: 13, & Use: 59 & Use: 105, \\
Domain: Well-being, & Domain: Innovation and Research, & Domain: Transport and Logistics, \\
Purpose: Mood tracking, & Purpose: Data collection and analysis, & Purpose: Facilitating contactless ticketing systems, \\
Capability: Analysing facial expressions for mood assessment, & Capability: Collecting and analysing facial data for research, & Capability: Recognising commuter faces for ticket validation, \\
AI User: Well-being apps, users, & AI User: Researchers, scientists, & AI User: Transport companies, \\
AI Subject: App users & AI Subject: Research subjects & AI Subject: Commuters \\
{\dotfill} & {\dotfill} & {\dotfill} \\
Use: 14, & Use: 60, & Use: 106, \\
Domain: Well-being, & Domain: Innovation and Research, & Domain: Urban Planning, \\
Purpose: Stress detection, & Purpose: Testing and improving facial recognition algorithms, & Purpose: Monitoring pedestrian traffic for city planning, \\
Capability: Identifying signs of stress on faces, & Capability: Using diverse facial data to refine algorithms, & Capability: Counting and tracking faces in public spaces, \\
AI User: Well-being apps, users, & AI User: Researchers, AI developers, & AI User: Urban planners, \\
AI Subject: App users & AI Subject: People in facial data sets & AI Subject: City residents \\
{\dotfill} & {\dotfill} & {\dotfill} \\
Use: 15, & Use: 61, & Use: 107, \\
Domain: Well-being, & Domain: Management and Operation of critical infrastructure, & Domain: Urban Planning, \\
Purpose: Personal growth coaching, & Purpose: Access control, & Purpose: Enhancing public safety in urban areas, \\
Capability: Analysing facial responses to personal growth exercises, & Capability: Verifying identities for secure access to facilities, & Capability: Identifying suspicious individuals in public spaces, \\
AI User: Personal growth apps, coaches, & AI User: Facility managers, security personnel, & AI User: City authorities, \\
AI Subject: Coaching clients & AI Subject: Employees, visitors & AI Subject: City residents \\
{\dotfill} & {\dotfill} & {\dotfill} \\
Use: 16, & Use: 62, & Use: 108, \\
Domain: Human-Computer Interaction, & Domain: Management and Operation of critical infrastructure, & Domain: Urban Planning, \\
Purpose: User authentication, & Purpose: Surveillance and security, & Purpose: Managing access to restricted urban areas, \\
Capability: Verifying user identity for system access, & Capability: Monitoring areas for unauthorised individuals, & Capability: Verifying authorised individuals for access, \\
AI User: Software developers, users, & AI User: Security personnel, facility managers, & AI User: City authorities, \\
AI Subject: Software users & AI Subject: People in monitored areas & AI Subject: City residents \\
{\dotfill} & {\dotfill} & {\dotfill} \\
Use: 17, & Use: 63, & Use: 109, \\
Domain: Human-Computer Interaction, & Domain: Management and Operation of critical infrastructure, & Domain: Counterterrorism, \\
Purpose: User experience personalisation, & Purpose: Emergency response, & Purpose: Identifying potential threats in public spaces, \\
Capability: Adapting system behaviour based on user's facial expressions, & Capability: Identifying individuals in emergency situations, & Capability: Recognising faces of individuals on watchlists, \\
AI User: Software developers, users, & AI User: Emergency responders, security personnel, & AI User: Security agencies, \\
AI Subject: Software users & AI Subject: People in emergency situations & AI Subject: General public \\
{\dotfill} & {\dotfill} & {\dotfill} \\
Use: 18, & Use: 64, & Use: 110, \\
Domain: Human-Computer Interaction, & Domain: Law enforcement, & Domain: Counterterrorism, \\
Purpose: Accessibility enhancement, & Purpose: Suspect identification, & Purpose: Verifying identity of individuals at checkpoints, \\
Capability: Enabling system control through facial gestures, & Capability: Matching faces to criminal databases, & Capability: Comparing faces to ID documents, \\
AI User: Software developers, users, & AI User: Police, investigators, & AI User: Security forces, \\
AI Subject: Users with physical disabilities & AI Subject: Suspects, victims & AI Subject: Individuals at checkpoints \\
{\dotfill} & {\dotfill} & {\dotfill} \\
Use: 19, & Use: 65, & Use: 111, \\
Domain: Finance and Investment, & Domain: Law enforcement, & Domain: Counterterrorism, \\
Purpose: Customer identification, & Purpose: Crowd monitoring, & Purpose: Investigating terrorist activities, \\
Capability: Verifying customer identity for transactions, & Capability: Identifying individuals in large crowds, & Capability: Analysing faces in surveillance footage, \\
AI User: Banks, financial institutions, & AI User: Police, security personnel, & AI User: Investigation agencies, \\
AI Subject: Bank customers & AI Subject: People in crowds & AI Subject: Suspected individuals \\
{\dotfill} & {\dotfill} & {\dotfill} \\
Use: 20, & Use: 66, & Use: 112, \\
Domain: Finance and Investment, & Domain: Law enforcement, & Domain: Environment and Sustainability, \\
Purpose: Fraud prevention, & Purpose: Investigation assistance, & Purpose: Monitoring wildlife populations, \\
Capability: Detecting fraudulent activities through facial recognition, & Capability: Analysing facial data from surveillance footage, & Capability: Recognising individual animals in a species, \\
AI User: Banks, financial institutions, & AI User: Investigators, police, & AI User: Conservation organisations, \\
AI Subject: Bank customers & AI Subject: People in surveillance footage & AI Subject: Wildlife \\
{\dotfill} & {\dotfill} & {\dotfill} \\
Use: 21, & Use: 67, & Use: 113, \\
Domain: Finance and Investment, & Domain: Migration, Asylum and Border control management, & Domain: Environment and Sustainability, \\
Purpose: Investor sentiment analysis, & Purpose: Identity verification, & Purpose: Tracking illegal poaching activities, \\
Capability: Analysing facial expressions for market sentiment, & Capability: Matching faces to passport or ID photos, & Capability: Identifying faces of known poachers, \\
AI User: Investment firms, traders, & AI User: Border control officers, immigration officials, & AI User: Wildlife protection agencies, \\
AI Subject: Investors & AI Subject: Travellers, migrants & AI Subject: Suspected poachers \\
{\dotfill} & {\dotfill} & {\dotfill} \\
Use: 22, & Use: 68, & Use: 114, \\
Domain: Education and vocational training, & Domain: Migration, Asylum and Border control management, & Domain: Environment and Sustainability, \\
Purpose: Student attendance tracking, & Purpose: Security checks, & Purpose: Managing access to protected areas, \\
Capability: Identifying students for attendance records, & Capability: Identifying individuals on watchlists, & Capability: Verifying authorised individuals for access, \\
AI User: Teachers, educational institutions, & AI User: Border control officers, immigration officials, & AI User: Park authorities, \\
AI Subject: Students & AI Subject: Travellers, migrants & AI Subject: Visitors \\
{\dotfill} & {\dotfill} & {\dotfill} \\
Use: 23, & Use: 69, & Use: 115, \\
Domain: Education and vocational training, & Domain: Migration, Asylum and Border control management, & Domain: International Law Enforcement and Cooperation, \\
Purpose: Learning engagement assessment, & Purpose: Asylum application processing, & Purpose: Identifying international criminals, \\
Capability: Analysing student facial expressions for engagement levels, & Capability: Verifying identities of asylum seekers, & Capability: Matching faces to international criminal databases, \\
AI User: Teachers, educational platforms, & AI User: Immigration officials, asylum officers, & AI User: Interpol, national law enforcement agencies, \\
AI Subject: Students & AI Subject: Asylum seekers & AI Subject: Suspected criminals \\
{\dotfill} & {\dotfill} & {\dotfill} \\
Use: 24, & Use: 70, & Use: 116, \\
Domain: Education and vocational training, & Domain: Democracy, & Domain: International Law Enforcement and Cooperation, \\
Purpose: Skill acquisition evaluation, & Purpose: Voter identification, & Purpose: Facilitating international prisoner transfers, \\
Capability: Assessing facial responses to vocational training tasks, & Capability: Verifying voter identities to prevent fraud, & Capability: Verifying identity of prisoners, \\
AI User: Trainers, vocational training institutions, & AI User: Election officials, poll workers, & AI User: Prison authorities, \\
AI Subject: Trainees & AI Subject: Voters & AI Subject: Prisoners \\
{\dotfill} & {\dotfill} & {\dotfill} \\
Use: 25, & Use: 71, & Use: 117, \\
Domain: Employment, workers management and access to self-employment, & Domain: Democracy, & Domain: International Law Enforcement and Cooperation, \\
Purpose: Employee attendance tracking, & Purpose: Public opinion analysis, & Purpose: Enhancing border security, \\
Capability: Recognising employee faces for timekeeping, & Capability: Analysing facial expressions in public gatherings, & Capability: Identifying individuals on watchlists at border crossings, \\
AI User: Human Resources, Management, & AI User: Political analysts, campaign managers, & AI User: Border control agencies, \\
AI Subject: Employees & AI Subject: People in public gatherings & AI Subject: Travellers \\
{\dotfill} & {\dotfill} & {\dotfill} \\
Use: 26, & Use: 72, & Use: 118, \\
Domain: Employment, workers management and access to self-employment, & Domain: Democracy, & Domain: Climate Change Mitigation and Adaptation, \\
Purpose: Access control to restricted areas, & Purpose: Public safety at political events, & Purpose: Monitoring deforestation activities, \\
Capability: Verifying employee identity for secure access, & Capability: Identifying potential threats in crowds, & Capability: Identifying individuals involved in illegal logging, \\
AI User: Security Personnel, Management, & AI User: Security personnel, event organisers, & AI User: Environmental agencies, \\
AI Subject: Employees & AI Subject: People at political events & AI Subject: Suspected illegal loggers \\
{\dotfill} & {\dotfill} & {\dotfill} \\
Use: 27, & Use: 73, & Use: 119, \\
Domain: Employment, workers management and access to self-employment, & Domain: Media and Communication, & Domain: Climate Change Mitigation and Adaptation, \\
Purpose: Remote worker identification, & Purpose: Enhancing content personalisation, & Purpose: Tracking carbon footprint of individuals, \\
Capability: Validating remote worker identity during virtual meetings, & Capability: Analysing viewer preferences and suggesting content, & Capability: Recognising individuals for carbon credit systems, \\
AI User: Management, Team Leaders, & AI User: Media platforms, content creators, & AI User: Climate change organisations, \\
AI Subject: Remote Employees & AI Subject: Media consumers & AI Subject: Individuals \\
{\dotfill} & {\dotfill} & {\dotfill} \\
Use: 28, & Use: 74, & Use: 120, \\
Domain: Essential private services and public services and benefits, & Domain: Media and Communication, & Domain: Climate Change Mitigation and Adaptation, \\
Purpose: Identity verification for service access, & Purpose: Verifying identity for secure communication, & Purpose: Managing access to climate-sensitive areas, \\
Capability: Confirming user identity for secure service access, & Capability: Authenticating users based on facial features, & Capability: Verifying authorised individuals for access, \\
AI User: Service Providers, Government Agencies, & AI User: Communication platforms, corporations, & AI User: Environmental authorities, \\
AI Subject: Service Users, Citizens & AI Subject: Communication platform users & AI Subject: Visitors \\
{\dotfill} & {\dotfill} & {\dotfill} \\
Use: 29, & Use: 75, & Use: 121, \\
Domain: Essential private services and public services and benefits, & Domain: Media and Communication, & Domain: Gaming and interactive experiences, \\
Purpose: Fraud prevention in public benefits, & Purpose: Improving audience engagement, & Purpose: Enhancing player immersion, \\
Capability: Detecting identity fraud in benefit claims, & Capability: Analysing audience reactions to content, & Capability: Translating player's facial expressions into game, \\
AI User: Government Agencies, & AI User: Advertisers, marketers, & AI User: Game developers, VR platforms, \\
AI Subject: Benefit Claimants & AI Subject: Audience members & AI Subject: Gamers \\
{\dotfill} & {\dotfill} & {\dotfill} \\
Use: 30, & Use: 76, & Use: 122, \\
Domain: Essential private services and public services and benefits, & Domain: Accessibility and Inclusion, & Domain: Gaming and interactive experiences, \\
Purpose: Automated passport control, & Purpose: Assisting visually impaired individuals, & Purpose: Improving game accessibility, \\
Capability: Verifying traveller identity at border controls, & Capability: Identifying faces and providing audio descriptions, & Capability: Enabling control through facial movements, \\
AI User: Border Control Agencies, & AI User: Accessibility software developers, & AI User: Game developers, accessibility designers, \\
AI Subject: Travellers & AI Subject: Visually impaired individuals & AI Subject: Disabled gamers \\
{\dotfill} & {\dotfill} & {\dotfill} \\
Use: 31, & Use: 77, & Use: 123, \\
Domain: Recommender Systems and Personalisation, & Domain: Accessibility and Inclusion, & Domain: Gaming and interactive experiences, \\
Purpose: Personalised advertising, & Purpose: Facilitating non-verbal communication, & Purpose: Creating personalised avatars, \\
Capability: Identifying user preferences for targeted ads, & Capability: Interpreting facial expressions and gestures, & Capability: Generating avatars based on player's face, \\
AI User: Advertisers, Online Platforms, & AI User: Communication app developers, & AI User: Game developers, social platforms, \\
AI Subject: Online Users & AI Subject: Non-verbal individuals & AI Subject: Gamers, social media users \\
{\dotfill} & {\dotfill} & {\dotfill} \\
Use: 32, & Use: 78, & Use: 124, \\
Domain: Recommender Systems and Personalisation, & Domain: Accessibility and Inclusion, & Domain: Hobbies, \\
Purpose: Content recommendation, & Purpose: Enhancing user interface accessibility, & Purpose: Enhancing photography, \\
Capability: Analysing user behaviour for personalised content, & Capability: Navigating software through facial movements, & Capability: Automatically focusing on faces in photos, \\
AI User: Streaming Platforms, Online Retailers, & AI User: Software developers, tech companies, & AI User: Photographers, camera manufacturers, \\
AI Subject: Consumers & AI Subject: Users with mobility impairments & AI Subject: Photography enthusiasts \\
{\dotfill} & {\dotfill} & {\dotfill} \\
Use: 33, & Use: 79, & Use: 125, \\
Domain: Recommender Systems and Personalisation, & Domain: Energy, & Domain: Hobbies, \\
Purpose: Personalised shopping experience, & Purpose: Monitoring energy consumption, & Purpose: Improving bird watching, \\
Capability: Recognising user for tailored shopping suggestions, & Capability: Identifying users and adjusting energy usage, & Capability: Identifying bird species from facial features, \\
AI User: Retailers, E-commerce Platforms, & AI User: Energy companies, smart home providers, & AI User: Bird watchers, app developers, \\
AI Subject: Shoppers & AI Subject: Homeowners, tenants & AI Subject: Bird watching enthusiasts \\
{\dotfill} & {\dotfill} & {\dotfill} \\
Use: 34, & Use: 80, & Use: 126, \\
Domain: Social Media, & Domain: Energy, & Domain: Hobbies, \\
Purpose: Photo tagging, & Purpose: Securing energy infrastructure, & Purpose: Personalising music experience, \\
Capability: Identifying individuals in photos for tagging, & Capability: Authenticating personnel access to facilities, & Capability: Adjusting music based on listener's expression, \\
AI User: Social Media Platforms, & AI User: Energy companies, security firms, & AI User: Music lovers, app developers, \\
AI Subject: Social Media Users & AI Subject: Energy facility personnel & AI Subject: Music enthusiasts \\
{\dotfill} & {\dotfill} & {\dotfill} \\
Use: 35, & Use: 81, & Use: 127, \\
Domain: Social Media, & Domain: Energy, & Domain: Smart home, \\
Purpose: Profile verification, & Purpose: Optimising energy distribution, & Purpose: Enhancing home security, \\
Capability: Verifying user identity to prevent fake profiles, & Capability: Identifying usage patterns and adjusting distribution, & Capability: Recognising authorised individuals for access, \\
AI User: Social Media Platforms, & AI User: Energy companies, grid operators, & AI User: Homeowners, security companies, \\
AI Subject: Social Media Users & AI Subject: Energy consumers & AI Subject: Home residents \\
{\dotfill} & {\dotfill} & {\dotfill} \\
Use: 36, & Use: 82, & Use: 128, \\
Domain: Social Media, & Domain: Military and Defence, & Domain: Smart home, \\
Purpose: Content moderation, & Purpose: Enhancing surveillance capabilities, & Purpose: Personalising user experience, \\
Capability: Detecting inappropriate or offensive images, & Capability: Identifying individuals in surveillance footage, & Capability: Adjusting settings based on user's presence, \\
AI User: Social Media Platforms, & AI User: Military, intelligence agencies, & AI User: Homeowners, smart device manufacturers, \\
AI Subject: Social Media Users & AI Subject: Surveillance targets & AI Subject: Home residents \\
{\dotfill} & {\dotfill} & {\dotfill} \\
Use: 37, & Use: 83, & Use: 129, \\
Domain: Sports and Recreation, & Domain: Military and Defence, & Domain: Smart home, \\
Purpose: Player identification, & Purpose: Improving personnel identification, & Purpose: Monitoring child safety, \\
Capability: Recognising players during live sports broadcasts, & Capability: Verifying identity at military installations, & Capability: Alerting when unrecognised faces are detected, \\
AI User: Broadcasters, Sports Leagues, & AI User: Military, defence contractors, & AI User: Parents, security companies, \\
AI Subject: Athletes, Viewers & AI Subject: Military personnel & AI Subject: Children \\
{\dotfill} & {\dotfill} & {\dotfill} \\
Use: 38, & Use: 84, & Use: 130, \\
Domain: Sports and Recreation, & Domain: Military and Defence, & Domain: Social and Community Services, \\
Purpose: Fan engagement, & Purpose: Facilitating threat assessment, & Purpose: Assisting in missing person cases, \\
Capability: Identifying fans for personalised experiences, & Capability: Identifying potential threats in crowds, & Capability: Matching faces in public footage to missing persons, \\
AI User: Sports Teams, Event Organisers, & AI User: Military, law enforcement agencies, & AI User: Law enforcement, social workers, \\
AI Subject: Sports Fans & AI Subject: Individuals in monitored areas & AI Subject: Missing persons \\
{\dotfill} & {\dotfill} & {\dotfill} \\
Use: 39, & Use: 85, & Use: 131, \\
Domain: Sports and Recreation, & Domain: Administration of justice and democratic processes, & Domain: Social and Community Services, \\
Purpose: Security at sports events, & Purpose: Assisting in criminal investigations, & Purpose: Enhancing public safety, \\
Capability: Identifying individuals for security purposes, & Capability: Identifying suspects in video footage, & Capability: Identifying individuals on watchlists in public spaces, \\
AI User: Event Security, Sports Leagues, & AI User: Law enforcement agencies, & AI User: Law enforcement, security agencies, \\
AI Subject: Event Attendees & AI Subject: Suspects, victims & AI Subject: General public \\
{\dotfill} & {\dotfill} & {\dotfill} \\
Use: 40, & Use: 86, & Use: 132, \\
Domain: Arts and Entertainment, & Domain: Administration of justice and democratic processes, & Domain: Social and Community Services, \\
Purpose: Audience analysis, & Purpose: Ensuring secure voting, & Purpose: Improving service accessibility, \\
Capability: Analysing audience reactions during performances, & Capability: Verifying voter identity at polling stations, & Capability: Facilitating sign language interpretation through facial expressions, \\
AI User: Performers, Event Organisers, & AI User: Election authorities, & AI User: Service providers, accessibility designers, \\
AI Subject: Audience Members & AI Subject: Voters & AI Subject: Deaf and hard of hearing individuals \\
{\dotfill} & {\dotfill} & {\dotfill} \\
Use: 41, & Use: 87, & Use: 133, \\
Domain: Arts and Entertainment, & Domain: Administration of justice and democratic processes, & Domain: Public and private transportation, \\
Purpose: Interactive exhibits, & Purpose: Facilitating courtroom identification, & Purpose: Enhancing passenger security, \\
Capability: Recognising visitors for interactive experiences, & Capability: Confirming identity of individuals in court proceedings, & Capability: Verifying passenger identity for boarding, \\
AI User: Museums, Art Galleries, & AI User: Courts, legal professionals, & AI User: Airlines, train companies, \\
AI Subject: Visitors & AI Subject: Defendants, witnesses & AI Subject: Passengers \\
{\dotfill} & {\dotfill} & {\dotfill} \\
Use: 42, & Use: 88, & Use: 134, \\
Domain: Arts and Entertainment, & Domain: Government Services and Administration, & Domain: Public and private transportation, \\
Purpose: Character creation in video games, & Purpose: Improving public service delivery, & Purpose: Improving driver safety, \\
Capability: Creating game characters based on user's face, & Capability: Identifying citizens for personalised services, & Capability: Detecting driver fatigue through facial analysis, \\
AI User: Game Developers, & AI User: Government agencies, & AI User: Car manufacturers, fleet managers, \\
AI Subject: Gamers & AI Subject: Citizens & AI Subject: Drivers \\
{\dotfill} & {\dotfill} & {\dotfill} \\
Use: 43, & Use: 89, & Use: 135, \\
Domain: Security and Cybersecurity, & Domain: Government Services and Administration, & Domain: Public and private transportation, \\
Purpose: Surveillance, & Purpose: Enhancing security at public facilities, & Purpose: Personalising in-vehicle experience, \\
Capability: Identifying individuals in surveillance footage, & Capability: Monitoring and identifying individuals at facilities, & Capability: Adjusting settings based on driver's preferences, \\
AI User: Law Enforcement, Security Companies, & AI User: Government agencies, security firms, & AI User: Car manufacturers, ride-sharing companies, \\
AI Subject: General Public & AI Subject: Public facility visitors & AI Subject: Drivers, passengers \\
{\dotfill} & {\dotfill} & {\dotfill} \\
Use: 44, & Use: 90, & Use: 136, \\
Domain: Security and Cybersecurity, & Domain: Government Services and Administration, & Domain: Interpersonal Communication, \\
Purpose: Access control, & Purpose: Facilitating document verification, & Purpose: Enhancing video communication, \\
Capability: Verifying identity for secure access, & Capability: Comparing facial features with ID photos, & Capability: Improving video quality by focusing on faces, \\
AI User: Security Personnel, IT Administrators, & AI User: Government agencies, & AI User: Video call platforms, users, \\
AI Subject: Employees, Users & AI Subject: Citizens, immigrants & AI Subject: Video call participants \\
{\dotfill} & {\dotfill} & {\dotfill} \\
Use: 45, & Use: 91, & Use: 137, \\
Domain: Security and Cybersecurity, & Domain: Diplomacy and Foreign Policy, & Domain: Interpersonal Communication, \\
Purpose: Identity verification in cybersecurity, & Purpose: Enhancing embassy security, & Purpose: Improving understanding of non-verbal cues, \\
Capability: Confirming user identity for secure online transactions, & Capability: Identifying individuals at diplomatic facilities, & Capability: Analyzing facial expressions during communication, \\
AI User: Cybersecurity Firms, Online Platforms, & AI User: Embassies, diplomatic security services, & AI User: Communication platforms, users, \\
AI Subject: Online Users & AI Subject: Embassy visitors, staff & AI Subject: Communication participants \\
{\dotfill} & {\dotfill} & {\dotfill} \\
Use: 46, & Use: 92, & Use: 138, \\
Domain: Marketing and Advertising, & Domain: Diplomacy and Foreign Policy, & Domain: Interpersonal Communication, \\
Purpose: Targeted advertising, & Purpose: Facilitating visa processing, & Purpose: Facilitating language learning, \\
Capability: Identifying user demographics for targeted ads, & Capability: Comparing applicant photos with passport photos, & Capability: Providing feedback on pronunciation through facial analysis, \\
AI User: Advertisers, Marketing Agencies, & AI User: Embassies, consulates, & AI User: Language learners, education platforms, \\
AI Subject: Consumers & AI Subject: Visa applicants & AI Subject: Language learners \\ \bottomrule
\end{longtable}

\subsection{(E) MATERIALS USED DURING USER STUDIES}

\begin{figure*}[!hbt]
\centering
\includegraphics[width=0.85\textwidth]{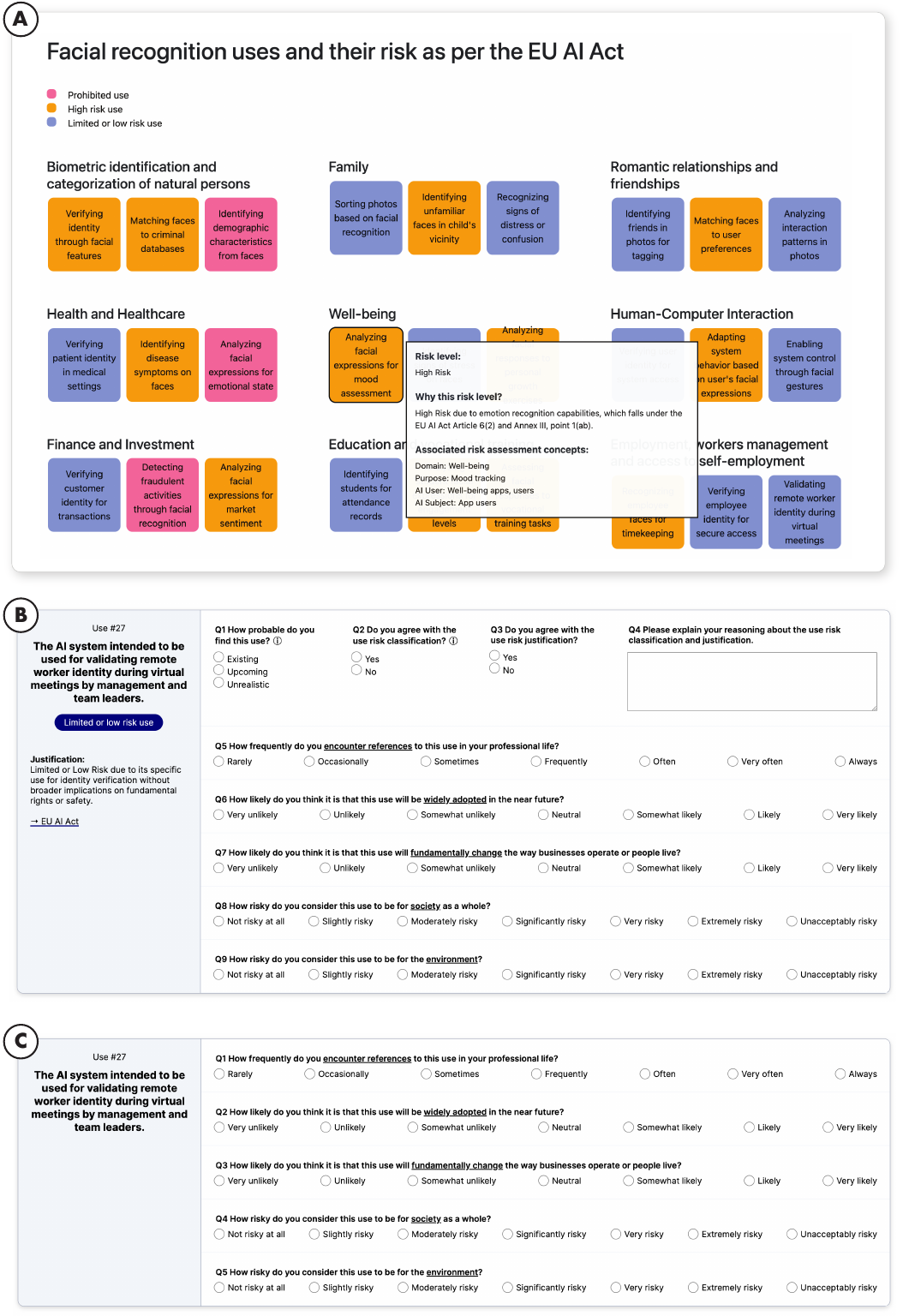}
\caption{\textbf{Materials used during user studies with AI practitioners}. 
During in-person studies, we showed AI developers and AI compliance experts an interactive list of 138 uses (A), followed by 16 interactive assessment cards for overlooked uses (B). 
During online studies, we showed AI developers and AI compliance experts a subset of 46 LLM-generated uses. In both in-person and online studies, AI developers interacted with a simplified version of the cards (B), while AI compliance experts used a more complex version (C), including the LLM-derived risk label, its justification, and questions to measure agreement between the experts and the LLM.}
\label{Fig:studies-materials}
\end{figure*}

\end{document}